\documentclass[a4paper,11pt]{article}
\usepackage{graphicx}
\usepackage{wrapfig}
\usepackage{amsmath}
\usepackage{amsfonts}
\usepackage{amssymb}
\usepackage{subcaption}
\usepackage{dsfont}
\usepackage{cite}
\usepackage{hyperref}
\usepackage{float}
\usepackage[title]{appendix}
\DeclareMathOperator{\csch}{csch}

\begin{document}
\title{\textbf{\boldmath Fulling-Davies-Unruh effect for accelerated 
two-level single
and entangled atomic systems}}
\author{
{\bf {\normalsize Arnab Mukherjee}\thanks{arnab.mukherjee@bose.res.in}}, 
{\bf {\normalsize Sunandan Gangopadhyay}
\thanks{ sunandan.gangopadhyay@bose.res.in}},
{\bf {\normalsize Archan S Majumdar}
\thanks{archan@bose.res.in}}\\
{\normalsize Department of Astrophysics and High Energy Physics},\\
{\normalsize S.N. Bose National Centre for Basic Sciences},\\
{\normalsize JD Block, Sector III, Salt Lake, Kolkata 700106, India}\\
}

\maketitle

\begin{abstract}
We investigate the transition rates of uniformly accelerated two-level single and  entangled atomic systems in empty space as well as inside a cavity.  
We take into account the interaction between the systems and a massless scalar field from the viewpoint of an instantaneously inertial observer and a coaccelerated observer, respectively.  The upward transition occurs only due to the acceleration of the atom. For the two-atom system, we consider that the system is initially prepared in a generic pure entangled state. In the presence of a cavity, we observe that for both the single and the two-atom cases, the upward and downward transitions are occurred due to the acceleration of the atomic systems. The transition rate manifests subtle features  depending upon the
cavity and system parameters, as well as the initial entanglement. It is
shown that no transition occurs for a maximally entangled super-radiant initial state, signifying that such entanglement in the accelerated two-atom system
can be preserved for quantum information procesing applications. 
Our analysis comprehensively validates  the equivalence between the effect of uniform acceleration for an inertial observer and the effect of a thermal bath
for a coaccelerated observer,  in free space as well as inside a cavity, if the temperature of the thermal bath is equal to the Unruh temperature.
\end{abstract}
	
\section{Introduction}\label{sec:Intro}

Relativistic quantum information is a growing area of study that combines ideas from gravitational physics with those  from quantum information theory \cite{Fuentes, Richter, Alsing, Bermudez, Hwang, Hwang1}. From the perspective of quantum communications, the fundamental role herein is
played by  quantum entanglement \cite{Haroche}. In recent times one of the key prototypes in research on entangled states in the relativistic domain are systems of two-level atoms interacting with quantum fields \cite{Plenio, Amico}. Radiative processes of entangled states have been extensively discussed in the literature  \cite{Agarwal}. In this regard, several important works
were developed \cite{Reznik2003, Reznik, Braun, Massar, Franson, Lin, Lin1}, which establish important results concerning entanglement generation between two localized causally disconnected
atoms. On the other hand, many investigations of atomic systems were also implemented on a curved background \cite{Menezes, Sen2022, Sen12022, Sen2023}.

Quantum field theory in curved background is another important area of theoretical physics that predicts observation is a frame dependent entity. As an example, one can consider that a uniformly accelerated particle detector sees the Minkowski vacuum as a thermal bath with temperature $T$ related to its proper acceleration $\alpha$, given by $T=\alpha/2\pi$. This phenomenon  arises as a result of the interaction between the detector and the fluctuating vacuum scalar fields, and is known as the Fulling-Davies-Unruh (FDU) effect \cite{Fulling, Davies1975, Unruh, birrell1984quantum}. After the seminal works of Fulling-Davies-Unruh \cite{Fulling, Davies1975, Unruh}, research into this phenomenon have been extended to include how a particle detector interacts with different quantum fields \cite{Frolov, Matsas3, OLevin, Muller1, Muller, Passante5, Matsas2, Crispino1, Hongwei1, Zhu3, Zhu2, Lin3, Zhu1, Zhou3, Zhu}. The application of both classical and  quantum field theory  has greatly improved the understanding of the origin of such phenomena \cite{Matsas3}. In addition to being significant, the FDU effect is also connected to a number of current research areas, including thermodynamics and the information paradox of a black hole \cite{Unruh, Crispino, Ross, Matsas1}.

The FDU effect has  been verified from different perspectives. It has 
been observed that when a uniformly accelerated single particle detector interacts with the vacuum massless scalar field, the average time variation of energy can be shown to be the same as seen by a locally inertial observer and by a coaccelerated observer. This equivalence holds both in free space as well as in presence of a reflecting boundary only if there exists a thermal bath at the FDU temperature in the coaccelerated frame \cite{Zhu1}. Investigations on the radiative properties of a single uniformly accelerated atom \cite{Hongwei1, Zhu3, Zhu2, Lin3, Zhu1, Zhou3} have also been extended to the scenarios where more than one atom in interaction with the massless scalar field and the electromagnetic field \cite{Passante4, Passante3, Passante2, Passante1, Passante, Menezes1, Menezes, Menezes2, Zhou2, Lima2019}.

Through the use of trapped atoms in optical nanofibers \cite{doi:10.1126/science.1237125, SOLANO2017439} and novel nanofabrication techniques \cite{Vetsch2010, Goban2012}, it is now possible to experimentally realize atomic excitations in nanoscale waveguides \cite{corzo2019waveguide}. The examination of fundamental quantum optical concepts like atom-photon lattices is made possible by these pathways \cite{RevModPhys.90.031002}. Studies on relativistic quantum phenomena in superconducting circuits \cite{PhysRevLett.110.113602, PhysRevB.92.064501} and secure quantum communication over long distances \cite{huang2019protection, huang2020deterministic, PhysRevD.104.105020, Aberg_2013} highlight the significance of reflecting boundaries. Reflecting boundaries also play an important physical role in the context of quantum entanglement \cite{zhang_2007, zhou2013boundary, Cheng_2018, liu2021entanglement}, holographic entanglement entropy \cite{akal2021holographic},  atom-field interaction \cite{Chatterjee2021_2},  and quantum thermodynamics \cite{Mukherjee2022}.

The basic motivation of investigating  the role of reflecting boundaries lies in its applicability to cavity quantum electrodynamics, a focus of fundamental research with numerous applications \cite{haroche2006exploring}. It has been observed that the resonance interatomic energy of two uniformly accelerated atoms can be effected due to the presence of boundaries \cite{Zhou2, Passante, Chatterjee2021_1} and noninertial atomic  motion \cite{Passante2}. To study the Unruh-Davies effect inside  cavities, techniques of cavity quantum electrodynamics   can be used \cite{Scully_2003, Scully_2006}. Additionally, cavity quantum electrodynamical configurations such as superconducting circuits \cite{PhysRevB.92.064501} and laser-driven technologies \cite{Bulanov2006, Eichmann2009, McWilliams2012} can achieve significant acceleration which is desirable for experimental verification of the theoretical results. Several theoretical analyses of the radiative processes of entangled atoms have been done by taking boundaries into account \cite{Scully_2003, Scully_2006, zhang_2007, zhou2013boundary, liu2021entanglement, arias2016boundary, Cheng_2018, Zhang2019}. 

In a recent work \cite{Zhou2020}, it has been found that there is an equivalence between the effect of uniform acceleration of an entangled two-atom system as observed by a Minkowski observer,  and that by a coaccelerated observer in free space  when the two-atom system is placed in a thermal bath. This equivalence only holds if the temperature of the thermal bath in the coaccelerated frame is taken to be equal to the Unruh temperature. On the 
other hand, the resonance interaction energy of a two-atom system as observed by an inertial observer and by a coaccelerated observer is  found to be 
the same in free space without considering any thermal bath at the Unruh temperature in the coaccelerated frame \cite{Passante1}. It has further been demonstrated  that the equivalence between the two-atom system uniformly accelerating with respect to the Minkowski observer and a static two-atom system (in free space) in a thermal bath breaks down \cite{Zhou2020}. 

The above results, with certain seemingly conflicting implications,
motivate us to perform a comprehensive investigation within the
same framework involving the status of the
FDU effect for both single and two-atomic entangled and accelerated systems 
in free space as well as in the presence of reflecting boundaries.
 Further motivation for our study in the context of cavities is two-fold. First of all, it is not clear    {\it a priori}, whether such an equivalence will still hold inside a cavity. The reason for this is the following. The physics inside a cavity is significantly different from that in  free space since a number of  field modes are curtailed due to boundary conditions. The second reason for carrying our investigation inside a cavity is that the cavity set-up is more realistic from an experimental point of view. Several
 recent experiments have been done using cavity set-up \cite{Lochan2020, Lochan2022, Arya2022, Vetsch2010, Goban2012}. 

In the present work we consider the interaction between the atomic systems and a massless scalar field in the frame of an instantaneously inertial observer and a coaccelerated observer, respectively.  The two-atom system is initially prepared in a generic pure entangled state. In the presence of a cavity, we 
show that for both the single and the two-atom cases, the magnitude of the upward and
downward transitions increase due  to the acceleration of the atomic systems. The transition rate displays interesting features  with variation of the 
cavity and system parameters, as well as the initial entanglement. We find
that no transition occurs for a maximally entangled super-radiant initial state, indicating that such entanglement in the accelerated two-atom system
can be preserved for quantum information procesing applications. 
We further compute values of the transition rate for two examples
using realistic cavity and system parameters.
From our analysis it follows that  the equivalence between the effect of uniform acceleration for an inertial observer and the effect of a thermal bath
for a coaccelerated observer, holds in free space as well as inside a cavity, if the temperature of the thermal bath is set equal to the Unruh temperature.

The paper is organized as follows: In section \ref{sec:CoupS}, we recapitulate
the basic framework  for obtaining the transition rate  when a single accelerated atom interacts with a massless scalar field. In section \ref{sec:Trate_inS}, we calculate the transition rates of the single atom from the viewpoint of the instantaneously inertial observer for empty space and in the presence of a cavity, respectively. A similar calculation of the transition rates of the single atom from the viewpoint of the coaccelerated observer for empty space and in the presence of a cavity, respectively ,is presented in section \ref{sec:Trate_coS}. We next consider the case of an
entangled and accelerated two-atom system from section \ref{sec:CoupT} onwards. In section \ref{sec:Trate_in}, we
study this system from the point of view of an inertial observer. Subsequently, in section \ref{sec:Trate_co} we calculate the tansition rate in context of the above system in context of a co-accelerated observer. We 
present a summary of our obtained results in section \ref{sec:Con}. Throughout the paper, we take $\hbar=c=k_B=1$, where $k_B$ is the Boltzmann constant.
\section{Coupling of a single atom with a massless scalar field}\label{sec:CoupS}
Let us consider a single atom with two energy levels, $-\omega_{0}/2$ and $+\omega_{0}/2$, travelling in  vacuum with massless scalar field fluctuations. In the laboratory frame, trajectories of the atom can be represented through $x(\tau)=(t(\tau),\mathbf{x}(\tau))$. In the instantaneous inertial frame, the Hamiltonian describing the atom-field interaction in the interaction picture is given by \cite{Svaiter1}
\begin{equation}
H = \lambda m(\tau) \phi(x(\tau))\,,\label{hamS}
\end{equation}
where $\lambda$ is the coupling constant which is assumed to be very small.
The mode expansion of the massless scalar field reads \cite{peskin2018introduction}
\begin{equation}
\phi(x(\tau))=\frac{1}{(2\pi)^{3/2}}\int_{-\infty}^{+\infty}\frac{d^{3}\mathbf{k}}{\sqrt{2 \omega_{\mathbf{k}}}}\Big[a_{\mathbf{k}}e^{-i\omega_{\mathbf{k}} t+i\mathbf{k}\cdot\mathbf{x}}+a_{\mathbf{k}}^{\dagger}e^{i\omega_{\mathbf{k}} t-i\mathbf{k}\cdot\mathbf{x}}\Big]\,\label{phiS}
\end{equation}
where $k=(\omega_\mathbf{k}, \mathbf{k})$ is the four momentum and $\mathbf{k}$ is the three momentum.
The monopole operator at any proper time $\tau$ of a single atom $m(\tau)$ is given by
\begin{equation}
m(\tau)=\exp{i H_{0}\tau}m(0)\exp{-i H_{0}\tau}\label{mt}
\end{equation}
with
$m(0)=\vert g\rangle\langle e\vert +\vert e\rangle\langle g\vert$
being the initial monopole operator and $H_0=\epsilon\vert e\rangle\langle e\vert$ being the free Hamiltonian of a single atom respectively \cite{Mukherjee2022}.

According to the time-dependent perturbation theory in the first-order approximation, the transition amplitude for the atom-field system from the initial atom-field state $\vert i\rangle\otimes\vert 0_M\rangle\equiv\vert i, 0_M\rangle$ to the final atom-field state $\vert f, \phi_{f}\rangle$ is
\begin{equation}
\mathcal{A}_{\vert i, 0_{M}\rangle\rightarrow\vert f, \phi_{f}\rangle}=i\lambda \langle f,\phi_{f}\vert \int_{-\infty}^{+\infty}m(\tau) \phi(x(\tau))\vert i,0_{M}\rangle\,\label{transampS}
\end{equation}
where $\vert i\rangle$ and $\vert f\rangle$ are the initial and final atomic states whereas $\vert 0_M\rangle$ and $\vert \phi_{f}\rangle$ are the initial (Minkowski vacuum state) and final field states.
Now, squaring the above transition amplitude and summing over all possible field states, transition probability from the initial state $\vert i\rangle$ to the final state $\vert f\rangle$ can be written as
\begin{equation}
\mathcal{P}_{\vert i\rangle\rightarrow \vert f\rangle}=\lambda^2 \vert m_{fi}\vert^2 F(\Delta E)\,,\label{transprobS}
\end{equation}
where $\Delta E= E_{f}-E_{i}$, $m_{fi}=\langle f\vert m(0)\vert i\rangle$ and
the response function $F(\Delta E)$ is defined as 
\begin{equation}
F(\Delta E)=\int_{-\infty}^{+\infty} d\tau \int_{-\infty}^{+\infty} d\tau'\,e^{-i\Delta E(\tau-\tau')}\,G^{+}(x(\tau),x(\tau'))\label{responseS}
\end{equation}
with  
\begin{equation}
G^{+}(x(\tau),x(\tau'))=\langle 0_{M}\vert\phi(x(\tau))\phi(x(\tau'))\vert 0_{M}\rangle\label{wightmanS}
\end{equation}
being the positive frequency Wightman function of the massless scalar field \cite{birrell1984quantum}.
Exploiting the time translational invariance property of the positive frequency Wightman function, the response function per unit proper time can be written as
\begin{equation}
\mathcal{F}(\Delta E)=\int_{-\infty}^{+\infty} d(\Delta\tau) \,e^{-i\Delta E\Delta\tau}\,G^{+}(x(\tau),x(\tau'))\label{responserateS}
\end{equation}
where $\Delta\tau=\tau-\tau'$.
Therefore, transition probability per unit proper time from the initial state $\vert i\rangle$ to the final state $\vert f\rangle$ turns out to be
\begin{equation}
\mathcal{R}_{\vert i\rangle\rightarrow \vert f \rangle}=\lambda^2 \vert m_{fi}\vert^2 \mathcal{F}(\Delta E)\,.\label{transprobrateS}
\end{equation}
In the following sections, the above formalism is used to examine the rate of transitions of a single atom  under various conditions such as non-inertial motion of the atom, nature of the observer, type of the background field and the presence of a cavity.
\section{Transition rates of a single atom from the viewpoint of a local inertial observer}\label{sec:Trate_inS}
In this section, we study the transitions of a uniformly accelerated single atom interacting with a massless scalar field from the perspective of a locally inertial observer. To see the boundary effects on the transitions of the uniformly accelerated single atom in this scenario, several cases have been studied in the following subsections.

\subsection{Transition rates in empty space with respect to a local inertial observer}

We first evaluate the transition rates of a single atom that has been uniformly accelerated while interacting with a vacuum massless scalar field in the absence of any perfectly reflecting boundary. In the laboratory frame, the atomic trajectory is given by
\begin{equation}
t(\tau)=\frac{1}{\alpha}\sinh(\alpha\tau),\,\,\,x(\tau)=\frac{1}{\alpha}\cosh(\alpha\tau),\,\,\,y=z=0\,,\label{trajecS}
\end{equation}
where $\alpha$ and $\tau$ denote the proper acceleration and the proper time of the atom.
Using the scalar field operator eq. \eqref{phiS} in eq. \eqref{wightmanS}, the Wightman function becomes \cite{birrell1984quantum}
\begin{align}
&G^{+}(x(\tau),x(\tau'))\nonumber\\
=&-\frac{1}{4\pi^2}\frac{1}{(t(\tau)-t(\tau')-i\varepsilon)^2-(x(\tau)-x(\tau'))^2-(y(\tau)-y(\tau'))^2-(z(\tau)-z(\tau'))^2}\,,\label{wightmn1S}
\end{align}
where $\varepsilon$ is a small positive number.
Substituting \eqref{trajecS} in \eqref{wightmn1S}), the Wightman function turns out to be
\begin{equation}
 G^{+}(x(\tau),x(\tau'))=-\frac{\alpha^2}{16\pi^2}\frac{1}{\sinh^2\left[\frac{1}{2}(\alpha\Delta\tau-i\varepsilon)\right]}\,. \label{Wght1S}
\end{equation}
Substituting the Wightman function into eq.\eqref{responserateS} and eq.\eqref{transprobrateS}, the transition rate from the initial state $\vert i\rangle$ to the final state $\vert f\rangle$ becomes 
\begin{equation}
\mathcal{R}_{\vert i\rangle\rightarrow \vert f\rangle}=-\frac{\lambda^2\vert m_{fi}\vert^2\alpha^2}{16\pi^2}\int_{-\infty}^{+\infty} d(\Delta\tau) \,e^{-i\Delta E\Delta\tau}\frac{1}{\sinh^2\left[\frac{1}{2}(\alpha\Delta\tau-i\varepsilon)\right]}\,. \label{transprobrate1S}
\end{equation}
 Simplifying the transition rates, eq.\eqref{transprobrate1S}, by performing the contour integration \cite{freitag2009complex} as shown in Appendix \ref{Appendix:A}, we obtain
 \begin{equation}
\mathcal{R}_{\vert i\rangle\rightarrow \vert f\rangle}=\frac{\lambda^2\vert m_{fi}\vert^2\vert \Delta E\vert}{2\pi}\left[\theta\,(-\Delta E)\left(1+\frac{1}{\exp(2\pi\vert \Delta E\vert/\alpha)-1}\right)+\theta\,(\Delta E)\left(\frac{1}{\exp(2\pi\Delta E/\alpha)-1}\right)\right] \label{transrateES}
\end{equation}
where $\theta(\Delta E)$ is the Heaviside step function defined as
\begin{equation}\label{delE}
\theta(\Delta E)=
\begin{cases}
1,\,\,\,\Delta E>0,\\
0,\,\,\,\Delta E<0.
\end{cases}
\end{equation}
The above equation eq.\eqref{transrateES} reveals that two transition processes, namely upward and downward transition can take place when the atom is under uniform acceleration. Considering the initial state $\vert i\rangle=\vert g\rangle$, final state $\vert f\rangle=\vert e\rangle$ and vice-versa, and using the definition $m_{eg}=\langle e\vert m(0)\vert g\rangle$, we obtain $\vert m_{ge}\vert^2=\vert m_{eg}\vert^2=1$, and $\Delta E=\omega_0$ for the transition $g\rightarrow e$ and $\Delta E=-\,\omega_0$ for the transition $e\rightarrow g$, respectively. Therefore, using the above results the upward and downward transition rates take the form 
\begin{equation}\label{up_empS}
\mathcal{R}_{\vert g\rangle\rightarrow \vert e\rangle}=\frac{\lambda^2\omega_0}{2\pi}\left(\frac{1}{\exp(2\pi\omega_0/\alpha)-1}\right)
\end{equation}
\begin{equation}\label{dwn_empS}
\mathcal{R}_{\vert e\rangle\rightarrow \vert g\rangle}=\frac{\lambda^2\omega_0}{2\pi}\left(1+\frac{1}{\exp(2\pi\omega_0/\alpha)-1}\right)\,.
\end{equation} 
The upward transition in free space occurs solely due to the acceleration
of the atom and vanishes in the limit $\alpha\rightarrow 0$. Taking the ratio of the above two results, we get
\begin{equation}\label{Ratio}
\frac{\mathcal{R}_{\vert g\rangle\rightarrow \vert e\rangle}}{\mathcal{R}_{\vert e\rangle\rightarrow \vert g\rangle}}\equiv\frac{\mathcal{R}_{up}}{\mathcal{R}_{down}}=\exp{-\frac{2\pi\omega_0}{\alpha}}\,.
\end{equation}
From the above expression, it is seen that the ratio of the upward and the downward transition rates depend only on the atomic acceleration and in the limit $\alpha\rightarrow\infty$, the ratio $\exp{-\frac{2\pi\omega_0}{\alpha}}\rightarrow1$, and hence, the two transition rates are equal in this limit.

\subsection{Transition rates in a cavity with respect to a local inertial observer}
\begin{figure}[H]
\centering
\includegraphics[scale=0.4]{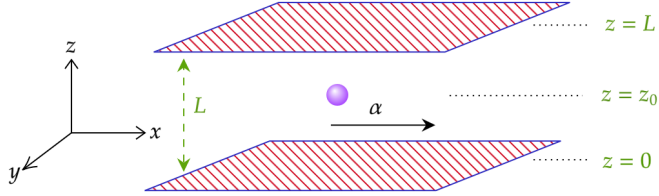}
\caption{Uniformly accelerated atom confined to a cavity.}\label{fig:SingleC_new}
\end{figure}

We now consider that a uniformly accelerated atom interacts with a vacuum massless scalar field confined to a cavity having length $L$ (see Figure \ref{fig:SingleC_new}) Assuming the scalar field obeys the Dirichlet boundary condition $\phi\vert_{z=0}=\phi\vert_{z=L}=0$, and using the method of images, the positive frequency Wightman function of the vacuum massless scalar field confined to the cavity of length $L$ takes the form \cite{birrell1984quantum}
\begin{align}
&G^{+}(x(\tau),x(\tau'))\nonumber\\
=&-\frac{1}{4\pi^2}\displaystyle\sum_{n=-\infty}^{\infty}\left[\frac{1}{(t(\tau)-t(\tau')-i\varepsilon)^2-(x(\tau)-x(\tau'))^2-(y(\tau)-y(\tau'))^2-(z(\tau)-z(\tau')-nL)^2}\right.\nonumber\\
&\left.-\frac{1}{(t(\tau)-t(\tau')-i\varepsilon)^2-(x(\tau)-x(\tau'))^2-(y(\tau)-y(\tau'))^2-(z(\tau)+z(\tau')-nL)^2}\right] \label{WghtCS}
\end{align}
with $\varepsilon$ is a small positive number.
 To represent the atomic trajectories in terms of the atomic proper time $\tau$, we choose the Cartesian coordinates in the laboratory frame so that the boundaries are fixed at $z=0$ and $z=L$.
Inside the cavity the atomic trajectory is given by
\begin{equation}
t(\tau)=\frac{1}{\alpha}\sinh(\alpha\tau),\,\,\,x(\tau)=\frac{1}{\alpha}\cosh(\alpha\tau),\,\,\,y=0,\,\,z=z_{0}\,.\label{trajecCS}
\end{equation}
Using the above trajectories in eq.\eqref{WghtCS}, the Wightman function becomes
\begin{equation}
 G^{+}(x(\tau),x(\tau'))=-\frac{\alpha^2}{16\pi^2}\displaystyle\sum_{n=-\infty}^{\infty}\left[\frac{1}{\sinh^2\left[\frac{1}{2}(\alpha\Delta\tau-i\varepsilon)\right]-\frac{1}{4}d_{1}^2\alpha^2}-\frac{1}{\sinh^2\left[\frac{1}{2}(\alpha\Delta\tau-i\varepsilon)\right]-\frac{1}{4}d_{2}^2\alpha^2}\right] \label{WghtC1S}
\end{equation}
with $d_1=nL,\,d_2=2z_{0}-nL$.
Substituting the Wightman function, eq. (\ref{WghtC1S}) into eq.\eqref{responserateS}, the transition rate from the initial state $\vert i\rangle$ to the final state $\vert f\rangle$ is given by
\begin{align}
\mathcal{R}_{\vert i\rangle\rightarrow \vert f\rangle}=&-\frac{\lambda^2\vert m_{fi}\vert^2\alpha^2}{16\pi^2}\displaystyle\sum_{n=-\infty}^{\infty}\left[\int_{-\infty}^{+\infty} d(\Delta\tau) \,e^{-i\Delta E\Delta\tau}\frac{1}{\sinh^2\left[\frac{1}{2}(\alpha\Delta\tau-i\varepsilon)\right]-\frac{1}{4}d_{1}^2\alpha^2}\right.\,\nonumber\\
&-\left.\int_{-\infty}^{+\infty} d(\Delta\tau) \,e^{-i\Delta E\Delta\tau}\frac{1}{\sinh^2\left[\frac{1}{2}(\alpha\Delta\tau-i\varepsilon)\right]-\frac{1}{4}d_{2}^2\alpha^2}\right]\,. \label{transprobrateC1S}
\end{align}
Simplifying the above equation through a  contour integral as shown in the Appendix \ref{Appendix:A}, the rate of transition from the initial state $\vert i\rangle$ to the final state $\vert f\rangle$ can be written as
\begin{align}
\mathcal{R}_{\vert i\rangle\rightarrow \vert f\rangle}=&\lambda^2\vert m_{fi}\vert^2\left[\theta\,(-\Delta E)\left\{\frac{\vert\Delta E\vert}{2\pi}+\mathfrak{f}\left(\vert \Delta E\vert,\,\alpha,\,\frac{L}{2}\right)-\mathfrak{h}\left(\vert \Delta E\vert,\,\alpha,\,z_{0},\frac{L}{2}\right)\right\}\left(1+\frac{1}{\exp{\frac{2\pi\vert \Delta E\vert}{\alpha}}-1}\right)\right.\nonumber\\
&+\left.\theta\,(\Delta E)\left\{\frac{\Delta E}{2\pi}+\mathfrak{f}\left(\Delta E,\,\alpha,\,\frac{L}{2}\right)-\mathfrak{h}\left(\Delta E,\,\alpha,\,z_{0},\frac{L}{2}\right)\right\}\left(\frac{1}{\exp{\frac{2\pi\Delta E}{\alpha}}-1}\right)\right]\label{transrateCES}
\end{align}
where we have defined
\begin{align}
\mathfrak{f}\left(\Delta E,\,\alpha,\,\frac{L}{2}\right)&=2\displaystyle\sum_{n=1}^{\infty}\mathfrak{g}\left(\Delta E,\,\alpha,\,\frac{nL}{2}\right)\label{ffunc1S}\\
\mathfrak{h}\left(\Delta E,\,\alpha,\,z_0\,,\frac{L}{2}\right)&=\displaystyle\sum_{n=-\infty}^{\infty}\mathfrak{g}\left(\Delta E,\,\alpha,\,z_0-\frac{nL}{2}\right)\,\label{hfunc1S}
\end{align}
where $\mathfrak{g}\left(\Delta E,\,\alpha,\,z_0\right)$ is defined as
\begin{equation}
\mathfrak{g}\left(\Delta E,\,\alpha,\,z_0\right)=\frac{\sin(\frac{2\Delta E}{\alpha}\sinh^{-1}(\alpha z_0))}{4\pi z_0\sqrt{1+\alpha^2 z_0^2}}\,.\label{gfuncS}
\end{equation}
Hence, from the above result the upward and downward transition rates can be written as
\begin{equation}\label{up_CS}
\mathcal{R}_{\vert g\rangle\rightarrow \vert e\rangle}=\lambda^2\left[\left\{\frac{\omega_0}{2\pi}+\mathfrak{f}\left(\omega_0,\,\alpha,\,\frac{L}{2}\right)-\mathfrak{h}\left(\omega_0,\,\alpha,\,z_{0},\frac{L}{2}\right)\right\}\left(\frac{1}{\exp{\frac{2\pi\omega_0}{\alpha}}-1}\right)\right]
\end{equation}
\begin{equation}\label{down_CS}
\mathcal{R}_{\vert e\rangle\rightarrow \vert g\rangle}=\lambda^2\left[\left\{\frac{\omega_0}{2\pi}+\mathfrak{f}\left(\omega_0,\,\alpha,\,\frac{L}{2}\right)-\mathfrak{h}\left(\omega_0,\,\alpha,\,z_{0},\frac{L}{2}\right)\right\}\left(1+\frac{1}{\exp{\frac{2\pi\omega_0}{\alpha}}-1}\right)\right]\,.
\end{equation}
Note that the ratio of the upward and the downward transition rates in the cavity scenario is identical with the free space result (eq.\eqref{Ratio}).

Next, in order to describe the single boundary and free space cases, we take the limiting cases of the above expressions. Taking the limit $L\rightarrow\infty$, we find that in eq.(s)(\ref{up_CS}, \ref{down_CS}) only  the $n=0$ term  survives from the infinite summation terms, and one can effectively reduce the cavity scenario to a situation where only one reflecting boundary exists. Hence, using this limit, the upward and downward transition rates in the presence of a single reflecting boundary turn out to be 
\begin{equation}
\mathcal{R}_{\vert g\rangle\rightarrow \vert e\rangle}=\lambda^2\left[\left\{\frac{\omega_0}{2\pi}-\mathfrak{g}\left(\omega_0,\,\alpha,\,z_{0}\right)\right\}\left(\frac{1}{\exp{2\pi\omega_0/\alpha}-1}\right)\right]
\end{equation}
\begin{equation}
\mathcal{R}_{\vert e\rangle\rightarrow \vert g\rangle}=\lambda^2\left[\left\{\frac{\omega_0}{2\pi}-\mathfrak{g}\left(\omega_0,\,\alpha,\,z_{0}\right)\right\}\left(1+\frac{1}{\exp{2\pi\omega_0/\alpha}-1}\right)\right]\,.
\end{equation}
On the other hand, taking the limits $L\rightarrow\infty$ and $z_{0}\rightarrow\infty$ together, eq.(s)(\ref{up_CS}, \ref{down_CS}) lead to the expression for the upward and downward transition rates in the free space given by eq.(s)(\ref{up_empS}, \ref{dwn_empS}).

We now investigate the variation of the transition rate of a single two level atom (from its ground state energy level $\vert g\rangle$ to the excited state energy level $\vert e\rangle$) confined to a cavity with the length of the cavity ($L$), distance of the atom from the boundary ($z_0$), and the atomic acceleration ($\alpha$). The findings are plotted below, where all physical quantities are expressed in dimensionless units.
\begin{figure}[H]
\centering
\includegraphics[scale=0.6]{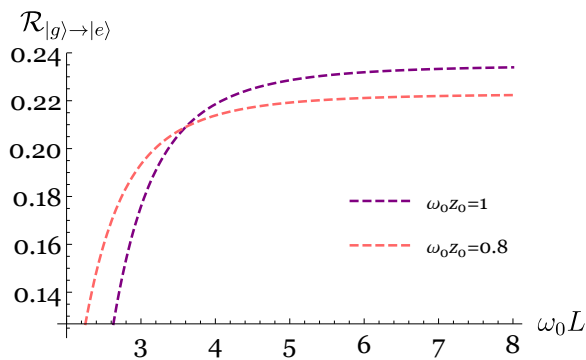}
\caption{Transition rate from $\vert g\rangle\rightarrow\vert e\rangle$ (per unit $\frac{\lambda^2\omega_0}{2\pi}$) versus separation between the two boundaries, $\alpha/\omega_0=4$.}\label{fig:PSingle_L}
\end{figure}
Figure \ref{fig:PSingle_L} shows the variation of the transition rate from $\vert g\rangle\rightarrow\vert e\rangle$ (per unit $\frac{\lambda^2\omega_0}{2\pi}$) with respect to the length of the cavity (separation between the two boundaries) for different values of distance of the atom from one boundary. From the plots, it can be seen that for a fixed value of the initial atomic distance $z_0$ from one boundary, the transition rate get enhanced when the cavity length increases and attains a maximum value for large values of $L$ ($\omega_0 L>>\omega_0 z_0$). 
This is expected since more number of field modes take part in the interaction between the scalar field and the atom after increasing the cavity length, which in turn increases the transition rate. When $\omega_0 L>>\omega_0 z_0$, the cavity scenario reduces to the case of a single boundary, and hence, the upward transition rate reaches a constant value. It is also observed that the 
upper value of the rate is more for a larger value of $\omega_0 z_0$.
\begin{figure}[H]
\centering
\includegraphics[scale=0.6]{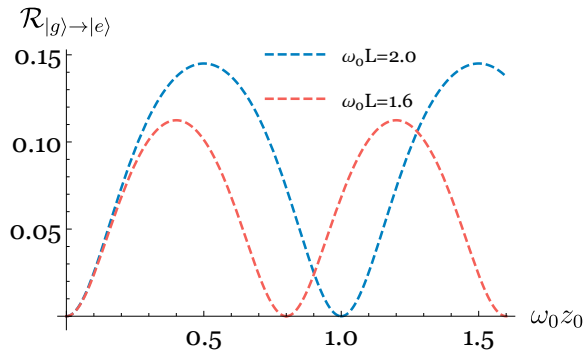}
\caption{Transition rate from $\vert g\rangle\rightarrow\vert e\rangle$ (per unit $\frac{\lambda^2\omega_0}{2\pi}$) versus distance of the atom from one boundary, $\alpha/\omega_0=4$.}\label{fig:PSingle_z}
\end{figure}
Figure \ref{fig:PSingle_z} shows the variation of the transition rate from $\vert g\rangle\rightarrow\vert e\rangle$ (per unit $\frac{\lambda^2\omega_0}{2\pi}$) with respect to the distance of the atom from one boundary for different values of the length of the cavity, for a fixed value of acceleration. From the plots, it is observed that for a fixed value of the length of the cavity $L$, when we increase the atomic distance from one boundary, transition rate shows an oscillatory behaviour and vanishes if either the atom touches any one of the boundaries or if the atom is equidistant from both boundaries. It can also be observed (as we have seen earlier in Fig \ref{fig:PSingle_L}) that with increase in the length of the cavity ($L$), the rate of upward transition increases.
\begin{figure}[H]
\centering
\includegraphics[scale=0.6]{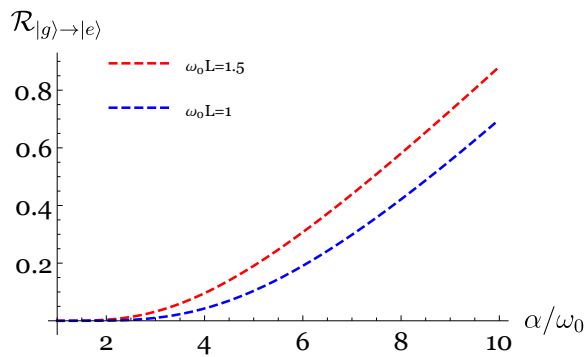}
\caption{Transition rate from $\vert g\rangle\rightarrow\vert e\rangle$ (per unit $\frac{\lambda^2\omega_0}{2\pi}$) versus acceleration, $\omega_0 z_0=0.3$.}\label{fig:PSingle_a}
\end{figure}
Figure \ref{fig:PSingle_a} shows the variation of the transition rate from $\vert g\rangle\rightarrow\vert e\rangle$ (per unit $\frac{\lambda^2\omega_0}{2\pi}$) with respect to the acceleration of the atom for different values of the length of the cavity and distance of the atom from one boundary. From the plots, it is observed that for a fixed value of the length of the cavity $L$ and the atomic distance $z_0$ from one boundary, the transition rate increases when the acceleration of the atom is increased. Once again we find that the transition rate is more for a larger value of the cavity length which is consistent with our earlier observations.
\begin{figure}[H]
\begin{minipage}{0.5\textwidth}
\centering
\includegraphics[scale=0.6]{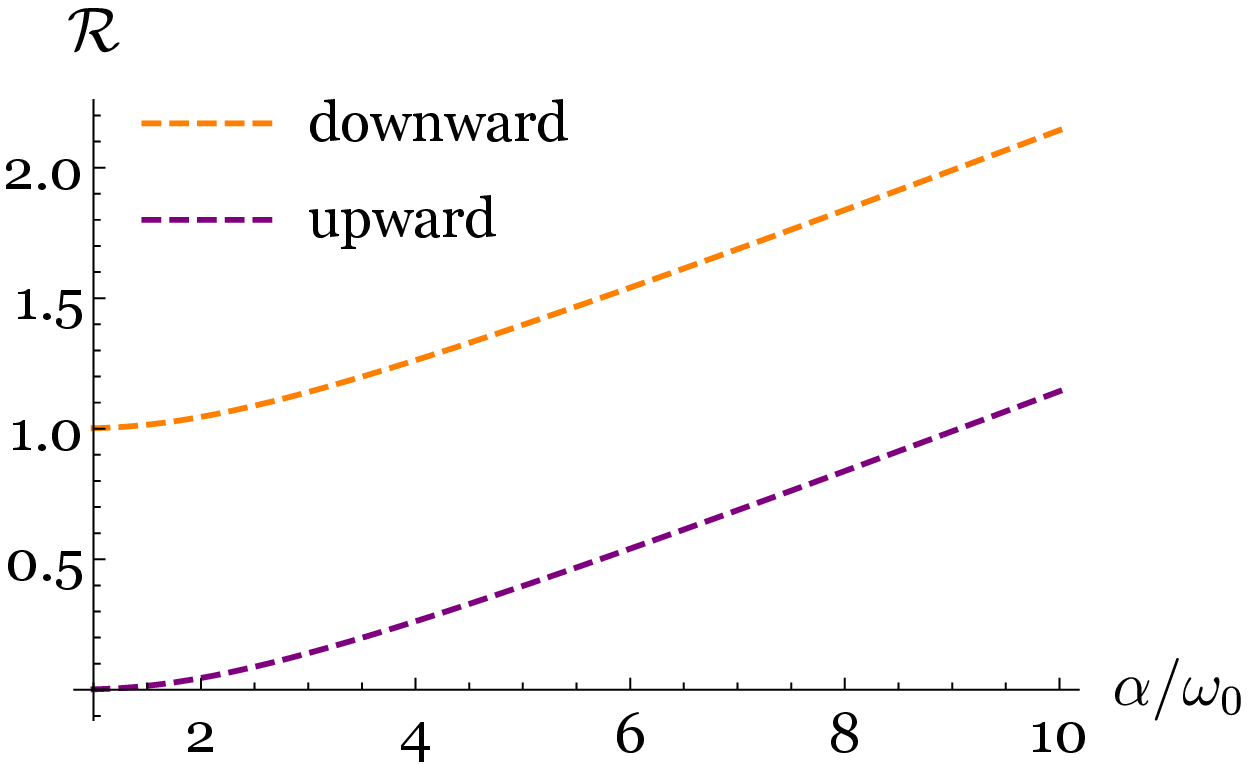}
\subcaption{In free space}\label{fig:RS_free}
\end{minipage}
\begin{minipage}{0.5\textwidth}
\centering
\includegraphics[scale=0.6]{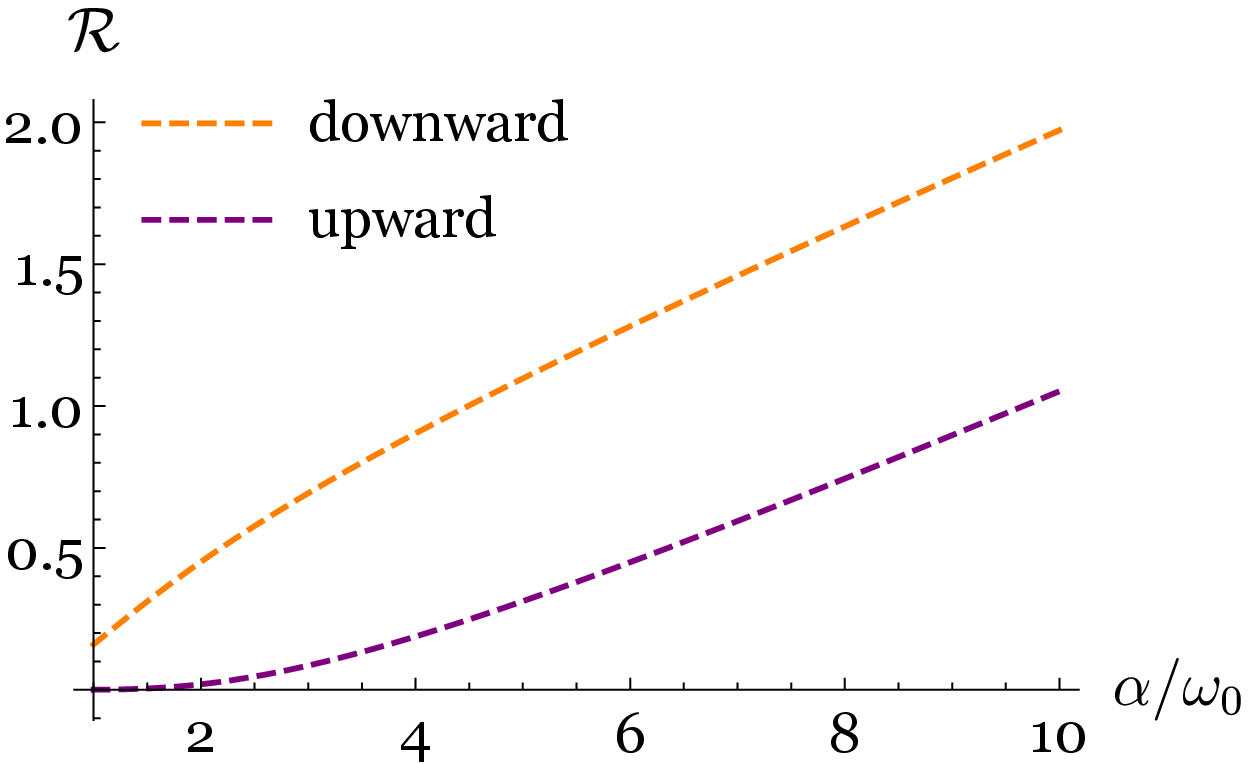}
\subcaption{Inside the cavity for a fixed value of $\omega_{0}L=3,\,\omega_{0}z_0=0.6$}\label{fig:RS_cavity}
\end{minipage}
\caption{Transition rate (per unit $\frac{\lambda^2\omega_0}{2\pi}$) versus acceleration.}\label{fig:RS_comp}
\end{figure}
In Figure \ref{fig:RS_comp}, we compare the upward and the downward transition rates with respect to the atomic acceleration for two cases, namely, atom in free space (Figure \ref{fig:RS_free}), atom confined to a cavity (Figure \ref{fig:RS_cavity}). From the above plots, it is seen that both the transition rates get affected due to the presence of the cavity. As noted
earlier, in presence of the cavity the upward and the downward transition rates decrease  with decrease in the cavity length. Here it is seen that when $\omega_0 L\sim\omega_0 z_0$, the cavity effect is strong enough to reduce
sharply the downward transition rate.

At this point, it might be interesting to make a quantitative estimation of the transition rate for the case when a single atom is placed inside a cavity. Following \cite{Chatterjee2021_1}, we choose the length of the cavity in the order of $100nm$, distance between the atom and the nearest boundary in the order of $20nm$, and the acceleration in the order of $10^{17}m/s^2$. The energy gap between the ground and excited state of a single Rubidium atom $Rb^{87}$ is of the order of $0.25eV$ \cite{Sansonetti}.  Now using eq.\eqref{up_CS} with $\lambda=0.1$, and the above values, the upward transition rate of the single atom inside a cavity turns out to be $3.38\times 10^{-12}eV=5.12\times10^{3} s^{-1}$. This tells us that in order to observe a transition  to the excited state from the ground state in $1 ns$, one would need to perform an experiment with a collection of $10^6$ atoms.
\section{Transition rates of a single atom from the viewpoint of a coaccelerated observer}\label{sec:Trate_coS}
In this section, the transitions of a uniformly accelerated atom is analysed from the perspective of a coaccelerated observer. The coaccelerated  observer  will perceive the atom as being static. Hence, the observer will see no Unruh acceleration radiation as there is no relative acceleration between the observer and the atom. However, for the observer to detect acceleration radiation, the field is assumed to be at an arbitrary temperature $T$. To calculate the transition rates and see the boundary effects on the transitions of the uniformly accelerated atom, we consider that the coordinates of the coaccelerated frame are the Rindler coordinates $(\tau,\,\eta,\,y,\,z)$ with the following relation with those of the laboratory coordinates ($t,\,x,\,y,\,z $)
\begin{equation}
t(\tau,\eta)=\frac{1}{\alpha}e^{\alpha\eta}\sinh(\alpha\tau),\,\,\,x(\tau,\eta)=\frac{1}{\alpha}e^{\alpha\eta}\cosh(\alpha\tau)\,.
\end{equation}
In the coaccelerated frame, the field operator $\phi(x(\tau))$ is replaced by its Rindler counterpart $\bar{\phi}(x(\tau))$ \cite{Zhou2020} and it takes the form \cite{Zhu1}
\begin{equation}
\bar{\phi}(\tau,\mathbf{x}))=\int_{0}^{\infty}d\omega\int_{-\infty}^{\infty}dk_{y}\int_{-\infty}^{\infty}dk_{z}\left[b_{\omega,k_{y},k_{z}}\mathcal{V}_{\omega,k_{y},k_{z}}(\tau,\mathbf{x})+b^{\dagger}_{\omega,k_{y},k_{z}}\mathcal{V}^{\star}_{\omega,k_{y},k_{z}}(\tau,\mathbf{x})\right]\label{phi_coS}
\end{equation}
with
\begin{equation}
\mathcal{V}_{\omega,k_{y},k_{z}}(\tau,\mathbf{x})=\sqrt{\frac{\sinh(\pi\omega/\alpha)}{4\pi^{4}\alpha}}\mathcal{K}_{i\frac{\omega}{\alpha}}\left(\frac{k_{\perp}}{\alpha}e^{\alpha\eta}\right)e^{-i\omega\tau+ik_{y}y+ik_{z}z}\label{modefnS}
\end{equation}
being the positive frequency orthonormal mode solution, $\mathcal{K}_{\nu}(x)$ is the Bessel function of imaginary argument and $k_{\perp}\equiv\vert\mathbf{k}_{\perp}\vert=\sqrt{k_{y}^2+k_{z}^2}$.
The interaction between the atom and the scalar field in this case can be written as \cite{Zhou2020}
\begin{equation}
H = \lambda m(\tau) \bar{\phi}(x(\tau))\,.\label{ham1S}
\end{equation}
As mentioned earlier, in order to determine the transition rate in the coaccelerated frame, we consider that the field is at an arbitrary temperature $T$. As the thermal state is a mixed state, therefore to calculate the response of a single atom coupled to the massless scalar field, it is further assumed that the field state can be represented by a pure state $\vert\sigma_{\omega,k_{y},k_{z}}\rangle$ with a probability factor $p_{\sigma}(\omega)=e^{-\beta\omega\sigma}/N(\omega)$ with $\beta=1/T$ and $N(\omega)=\displaystyle\sum_{\sigma=0}^{\infty}e^{-\beta\omega\sigma}$. In this case, $\vert \psi_{\pm},\,\sigma_{\omega,k_{y},k_{z}}\rangle$ and $\vert E_n,\,\gamma_{\omega',k^{\prime}_{y},k^{\prime}_{z}}\rangle$ can be used to represent respectively, the initial and the final state of the atom-field system.

Now, following the procedure described in the previous section, the 
transition probability of the atom-field system  from the initial state $\vert i\rangle$ to final state $\vert f\rangle$ is  given by 
\begin{equation}
\mathcal{P}^{\beta}_{\vert i\rangle\rightarrow \vert f\rangle}=\lambda^2 \vert m_{fi}\vert^2 F^{\beta}(\Delta E)\,,\label{transprob-coS}
\end{equation}
where the response function $F^{\beta}(\Delta E)$ is defined as 
\begin{equation}
F^{\beta}(\Delta E)=\int_{-\infty}^{+\infty} d\tau \int_{-\infty}^{+\infty} d\tau'\,e^{-i\Delta E(\tau-\tau')}\,G_{\beta}^{+}(x(\tau),x(\tau'))\label{response-coS}
\end{equation} 
and 
\begin{align}
&G^{+}_{\beta}(x(\tau),x(\tau'))=\frac{tr[\rho'\phi(x(\tau))\phi(x(\tau')]}{tr[\rho']}\nonumber\\
&=N^{-1}(\omega)\displaystyle\sum_{\sigma=0}^{\infty}\int_{0}^{\infty}d\omega\int_{-\infty}^{\infty}dk_{y}\int_{-\infty}^{\infty}dk_{z}e^{-\beta\omega\sigma}\langle \sigma_{\omega,k_{y},k_{z}}\vert\bar{\phi}(x(\tau))\bar{\phi}(x(\tau'))\vert \sigma_{\omega,k_{y},k_{z}}\rangle\label{wightman-coS}
\end{align}
is the positive frequency Wightman function of the scalar field in a thermal state at an arbitrary temperature $T$ in the coaccelerated frame.
Exploiting the time translational invariance property of the positive frequency Wightman function, the response function per unit proper time can be written as
\begin{equation}
\mathcal{F}^{\beta}(\Delta E)=\int_{-\infty}^{+\infty} d(\Delta\tau) \,e^{-i\Delta E\Delta\tau}\,G^{+}_{\beta}(x(\tau),x(\tau'))\,.\label{responserate-coS}
\end{equation}
Therefore, the transition probability per unit proper time of the atom from the initial state $\vert i\rangle$ to the final state $\vert f\rangle$ turns out to be
\begin{equation}
\mathcal{R}^{\beta}_{\vert i\rangle\rightarrow \vert f\rangle}=\lambda^2 \vert m_{fi}\vert^2 \mathcal{F}^{\beta}(\Delta E)\,.\label{transprobrate-coS}
\end{equation}
\subsection{Transition rates in empty space with respect to a coaccelerated observer}
Using eq.\eqref{phi_coS} in eq.\eqref{wightman-coS}, the thermal Wightman function takes the following form for an arbitrary temperature $T$,
\begin{align}
&G^{+}_{\beta}(x(\tau),x(\tau'))\nonumber\\
&=\int_{0}^{\infty}d\omega\int_{-\infty}^{\infty}dk_{y}\int_{-\infty}^{\infty}dk_{z}\bigg[\displaystyle\sum_{\sigma=0}^{\infty}(\sigma+1)e^{-\beta\omega\sigma}\mathcal{V}_{\omega,k_{y},k_{z}}(\tau,\mathbf{x})\mathcal{V}^{\star}_{\omega,k_{y},k_{z}}(\tau^{\prime},\mathbf{x}^{\prime})\bigg.\nonumber\\
&+\bigg.\displaystyle\sum_{\sigma=1}^{\infty}\sigma e^{-\beta\omega\sigma}\mathcal{V}^{\star}_{\omega,k_{y},k_{z}}(\tau,\mathbf{x})\mathcal{V}_{\omega,k_{y},k_{z}}(\tau^{\prime},\mathbf{x}^{\prime})\bigg]\bigg/\displaystyle\sum_{\sigma=0}^{\infty}e^{-\beta\omega\sigma}\nonumber\\
&=\int_{0}^{\infty}d\omega\int_{-\infty}^{\infty}dk_{y}\int_{-\infty}^{\infty}dk_{z}\bigg[\frac{e^{\omega/T}}{e^{\omega/T}-1}\mathcal{V}_{\omega,k_{y},k_{z}}(\tau,\mathbf{x})\mathcal{V}^{\star}_{\omega,k_{y},k_{z}}(\tau^{\prime},\mathbf{x}^{\prime})\bigg.\nonumber\\
&+\bigg.\frac{1}{e^{\omega/T}-1}\mathcal{V}^{\star}_{\omega,k_{y},k_{z}}(\tau,\mathbf{x})\mathcal{V}_{\omega,k_{y},k_{z}}(\tau^{\prime},\mathbf{x}^{\prime})\bigg]\,.\label{wightman-coS1}
\end{align}
In the Rindler coordinates, the trajectory of the atom can be described by
\begin{equation}
t=\tau,\,\,\,\eta=y=z=0\,.\label{trajec_coS}
\end{equation}
Using eq.(s)(\ref{modefnS}, \ref{trajec_coS}) in eq.\eqref{wightman-coS1}, the thermal Wightman function takes the form
\begin{align}
&G^{+}_{\beta}(x(\tau),x(\tau'))\nonumber\\
&=\frac{1}{4\pi^{4}\alpha}\int_{0}^{\infty}d\omega\int_{-\infty}^{\infty}dk_{y}\int_{-\infty}^{\infty}dk_{z}\sinh\left(\frac{\pi\omega}{\alpha}\right)\mathcal{K}_{i\omega/\alpha}^{2}\left(\frac{k_{\perp}}{\alpha}\right)\bigg[\frac{e^{\omega/T}}{e^{\omega/T}-1}e^{-i\omega(\tau-\tau')}\bigg.\nonumber\\
&+\bigg.\frac{1}{e^{\omega/T}-1}e^{i\omega(\tau-\tau')}\bigg]\\
&=\frac{1}{4\pi^{2}}\int_{0}^{\infty}d\omega\,\omega\bigg[\frac{e^{\omega/T}}{e^{\omega/T}-1}e^{-i\omega(\tau-\tau')}+\bigg.\frac{1}{e^{\omega/T}-1}e^{i\omega(\tau-\tau')}\bigg]\nonumber\\
&=-\frac{1}{4\pi^{2}}\displaystyle\sum_{s=-\infty}^{\infty}\frac{1}{(\Delta\tau-is\beta-i\varepsilon)^2}\label{wightman-coS2}
\end{align}
where in the second line, we have used the integral
\begin{equation}
\int_{-\infty}^{\infty}dk_{y}\int_{-\infty}^{\infty}dk_{z}\mathcal{K}_{i\omega/\alpha}^{2}\left(\frac{k_{\perp}}{\alpha}\right)=\frac{\alpha\pi^2\omega}{\sinh(\pi\omega/\alpha)}\,.
\end{equation}
By inserting the above Wightman function into eq.(s)(\ref{responserate-coS}, \ref{transprobrate-coS}), and performing integrations using the contour integration method, the upward and downward transition rates of a single atom submerged in the thermal bath turn out to be, respectively
\begin{equation}\label{upco_empS}
\mathcal{R}^{\beta}_{\vert g\rangle\rightarrow \vert e\rangle}=\frac{\lambda^2\omega_0}{2\pi}\left(\frac{1}{\exp(\omega_0/T)-1}\right)
\end{equation}
\begin{equation}\label{dwnco_empS}
\mathcal{R}^{\beta}_{\vert e\rangle\rightarrow \vert g\rangle}=\frac{\lambda^2\omega_0}{2\pi}\left(1+\frac{1}{\exp(\omega_0/T)-1}\right)\,.
\end{equation} 
The above equations suggest that in the coaccelerated frame both the upward and the downward transition can occur for an atom immersed in the thermal bath which is very similar to the transitions observed by an instantaneous inertial observer. Taking the limit $T\rightarrow 0$, we can see that the upward transition rate vanishes and this is consistent with the fact that there should be no transition if the observer is static with respect to the atom. Eq.(s)(\ref{up_empS}, \ref{dwn_empS}, \ref{upco_empS}, \ref{dwnco_empS}) clearly indicate that the transition rates of an uniformly accelerated atom seen by an instantaneously inertial observer and by a coaccelerated observer are identical only when if we take the thermal bath temperature in the coaccelerated frame to be $T=\alpha/2\pi$.
\subsection{Transition rates in a cavity from the viewpoint of a coaccelerated observer}
In this subsection, we consider a uniformly accelerated atom interacting with a vacuum massless scalar field confined to a cavity of length $L$ from the perspective of a coaccelerated observer. We assume that perfectly reflecting boundaries are placed at $z = 0$ and $z=L$. In a coaccelerated frame, this scenario will be depicted as a static atom interacting with a massless scalar field in a thermal state at an arbitrary temperature $T$ inside a cavity of length $L$.

In case of a single reflecting boundary, the scalar field obeys the Dirichlet boundary condition $ \phi\vert_{z=0}=0$. The positive frequency Rindler mode function for the massless scalar field takes the form
\begin{equation}
\mathcal{V}_{\omega,k_{y},k_{z}}(\tau,\mathbf{x})=\sqrt{\frac{\sinh(\pi\omega/\alpha)}{2\pi^{4}\alpha}}\mathcal{K}_{i\frac{\omega}{\alpha}}\left(\frac{k_{\perp}}{\alpha}e^{\alpha\eta}\right)\sin(k_{z}z)e^{-i\omega\tau+ik_{y}y}\,.\label{modefn1S}
\end{equation}
Inserting eq.\eqref{modefn1S} in eq.\eqref{wightman-coS1}, the thermal Wightman function takes the form
\begin{align}
&G^{+}_{\beta}(x(\tau),x(\tau'))\nonumber\\
&=\frac{1}{4\pi^{4}\alpha}\int_{0}^{\infty}d\omega\int_{-\infty}^{\infty}dk_{y}\int_{-\infty}^{\infty}dk_{z}\sinh\left(\frac{\pi\omega}{\alpha}\right)\mathcal{K}_{i\omega/\alpha}\left(\frac{k_{\perp}}{\alpha}e^{\alpha\eta}\right)\mathcal{K}_{i\omega/\alpha}\left(\frac{k_{\perp}}{\alpha}e^{\alpha\eta^{\prime}}\right)\bigg\{\cos\big[k_{z}(z-z^{\prime})\big]\bigg.\nonumber\\
&-\bigg.\cos\big[k_{z}(z+z^{\prime})\big]\bigg\}\times\bigg[\frac{e^{\omega/T}}{e^{\omega/T}-1}e^{-i\omega(\tau-\tau^{\prime})+ik_{y}(y-y^{\prime})}+\frac{1}{e^{\omega/T}-1}e^{i\omega(\tau-\tau^{\prime})-ik_{y}(y-y^{\prime})}\bigg]\,.\label{wightman-coBS}
\end{align}
Now for the cavity scenario, the Dirichlet boundary condition obeyed by the scalar field is $\phi\vert_{z=0}=\phi\vert_{z=L}=0$. Using the above boundary condition (eq.\eqref{wightman-coBS}) and by using the method of images, the thermal Wightman function of the massless scalar field confined to the cavity turns out to be
\begin{align}
&G^{+}_{\beta}(x(\tau),x(\tau'))\nonumber\\
&=\frac{1}{4\pi^{4}\alpha}\displaystyle\sum_{n=-\infty}^{\infty}\int_{0}^{\infty}d\omega\int_{-\infty}^{\infty}dk_{y}\int_{-\infty}^{\infty}dk_{z}\sinh\left(\frac{\pi\omega}{\alpha}\right)\mathcal{K}_{i\omega/\alpha}\left(\frac{k_{\perp}}{\alpha}e^{\alpha\eta}\right)\mathcal{K}_{i\omega/\alpha}\left(\frac{k_{\perp}}{\alpha}e^{\alpha\eta^{\prime}}\right)\nonumber\\
&\times\bigg\{\cos\big[k_{z}(z-z^{\prime}-nL)\big]-\cos\big[k_{z}(z+z^{\prime}-nL)\big]\bigg\}\bigg[\frac{e^{\omega/T}}{e^{\omega/T}-1}e^{-i\omega(\tau-\tau^{\prime})+ik_{y}(y-y^{\prime})}\bigg.\nonumber\\
&+\bigg.\frac{1}{e^{\omega/T}-1}e^{i\omega(\tau-\tau^{\prime})-ik_{y}(y-y^{\prime})}\bigg]\,.\label{wightman-coCS}
\end{align}

\begin{figure}[H]
\centering
\includegraphics[scale=0.4]{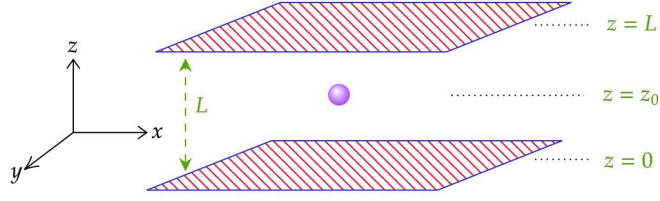}
\caption{Static atom confined to a cavity.}\label{fig:SingleC1_new}
\end{figure}
Inside the cavity, the atomic trajectory becomes
\begin{equation}
t=\tau,\,\,\,\eta=y=0,\,\,z=z_0\,.\label{trajec_coCS}
\end{equation}
Inserting the above trajectory in eq.\eqref{wightman-coCS}, using the result \cite{Zhu1}
\begin{equation}
\int_{-\infty}^{\infty}dk_{y}\int_{-\infty}^{\infty}dk_{z}\mathcal{K}_{i\omega/\alpha}^{2}\left(\frac{k_{\perp}}{\alpha}\right)\cos\left[2k_{z}\frac{d}{2}\right]=\frac{\alpha\pi^2}{\sinh(\pi\omega/\alpha)}\frac{\sin\left(\frac{2\omega}{\alpha}\sinh^{-1}\left(\frac{d\alpha}{2}\right)\right)}{d\sqrt{1+\frac{1}{4}\alpha^2 d^2}}
\end{equation}
and following the procedure mentioned in the previous subsection, the thermal Wightman function becomes
\begin{equation}
G^{+}_{\beta}(x(\tau),x(\tau'))=-\frac{1}{4\pi^{2}}\displaystyle\sum_{n=-\infty}^{\infty}\displaystyle\sum_{s=-\infty}^{\infty}\left[\frac{\mathcal{B}_1}{\mathcal{C}_1}\frac{1}{(\Delta\tau-is\beta-i\varepsilon)^2-\mathcal{B}_{1}^2}-\frac{\mathcal{B}_2}{\mathcal{C}_2}\frac{1}{(\Delta\tau-is\beta-i\varepsilon)^2-\mathcal{B}_{2}^2}\right]\label{wightman-coCS1}
\end{equation}
with 
\begin{align*}
\mathcal{B}_1&=\frac{2}{\alpha}\sinh^{-1}\left(\frac{nL\alpha}{2}\right)\\
\mathcal{C}_1&=nL\sqrt{1+\frac{1}{4}\alpha^2 n^2L^2}\\
\mathcal{B}_2&=\frac{2}{\alpha}\sinh^{-1}\left(\alpha\left(z_{0}-\frac{nL}{2}\right)\right)\\
\mathcal{C}_2&=(2z_{0}-nL)\sqrt{1+\frac{1}{4}\alpha^2 (2z_{0}-nL)^2}\,.
\end{align*}
Putting the above Wightman function in eq.(s)(\ref{responserate-coS}, \ref{transprobrate-coS}), and performing the integrations using the contour integration technique, the downward and upward transition rates of the single atom system submerged in the thermal bath read, respectively
\begin{equation}\label{upco_CS1}
\mathcal{R}_{\vert g\rangle\rightarrow \vert e\rangle}=\lambda^2\left[\left\{\frac{\omega_0}{2\pi}+\mathfrak{f}\left(\omega_0,\,\alpha,\,\frac{L}{2}\right)-\mathfrak{h}\left(\omega_0,\,\alpha,\,z_{0},\frac{L}{2}\right)\right\}\left(\frac{1}{\exp(\omega_0/T)-1}\right)\right]
\end{equation}
\begin{equation}\label{downco_CS1}
\mathcal{R}_{\vert e\rangle\rightarrow \vert g\rangle}=\lambda^2\left[\left\{\frac{\omega_0}{2\pi}+\mathfrak{f}\left(\omega_0,\,\alpha,\,\frac{L}{2}\right)-\mathfrak{h}\left(\omega_0,\,\alpha,\,z_{0},\frac{L}{2}\right)\right\}\left(1+\frac{1}{\exp(\omega_0/T)-1}\right)\right]\,.
\end{equation}
Next, taking the limit $L\rightarrow\infty$, the upward and the downward transition rate in the presence of a single reflecting boundary turn out 
respectively to be 
\begin{equation}
\mathcal{R}_{\vert g\rangle\rightarrow \vert e\rangle}=\lambda^2\left[\left\{\frac{\omega_0}{2\pi}-\mathfrak{g}\left(\omega_0,\,\alpha,\,z_{0}\right)\right\}\left(\frac{1}{\exp{\omega_0/T}-1}\right)\right]
\end{equation}
\begin{equation}
\mathcal{R}_{\vert e\rangle\rightarrow \vert g\rangle}=\lambda^2\left[\left\{\frac{\omega_0}{2\pi}-\mathfrak{g}\left(\omega_0,\,\alpha,\,z_{0}\right)\right\}\left(1+\frac{1}{\exp{\omega_0/T}-1}\right)\right]
\end{equation}
with $\mathfrak{f}\left(\omega_{0},\,\alpha,\,\frac{L}{2}\right)$, $\mathfrak{h}\left(\omega_{0},\,\alpha,\,z_{0},\,\frac{L}{2}\right)$ and $\mathfrak{g}\left(\omega_{0},\,\alpha,\,z_{0}\right)$ being the same as in eq.(s)(\ref{ffunc1S}, \ref{hfunc1S}) and eq.\eqref{gfuncS}.

The above analysis clearly displays the  similarity between the transitions observed by an instantaneously inertial observer and a coaccelerated observer in a thermal bath  for both the upward and the downward transition rates    when the atom is confined in a cavity. Here too we notice that taking the thermal bath temperature in the coaccelerated frame $T=\alpha/2\pi$, eq.(s)(\ref{upco_CS1}, \ref{downco_CS1}, \ref{up_CS}, \ref{down_CS}) indicate that the transition rates of a uniformly accelerated atom seen by a coaccelerated observer in a thermal bath and by an instantaneously inertial observer are identical inside the cavity.

\section{Coupling of the two-atom system with a massless scalar field}\label{sec:CoupT}
We consider two identical atoms $A$ and $B$ and assume that they are travelling along synchronous  trajectories in a vacuum with massless scalar field fluctuations, the interatomic distance is assumed to be constant and the proper times of the two atoms can be described by the same time $\tau$ \cite{Zhang2019}. In the laboratory frame, trajectories of the two atoms can be represented through $x_{A}(\tau)$ and $x_{B}(\tau)$. Here we consider each atom as a two level system having energy levels $-\omega_{0}/2$ and $+\omega_{0}/2$. Therefore, the entire two-atom system can be described by the three energy levels with energies $-\omega_{0},\,0\,\,\text{and}\,\,\omega_{0}$ \cite{Svaiter2}. We designate them by $E_n$ with $n=1,2,3$. The low and high energy levels associated with eigenstates are $\vert E_1\rangle=\vert g_{A},g_{B}\rangle$ and $\vert E_3\rangle=\vert e_{A},e_{B}\rangle$ where $\vert g\rangle$ and $\vert e\rangle$ represent the ground state and the excited state of a single atom respectively. The energy level $E_2$ is degenerate corresponding to the eigenstates $\vert g_{A},e_{B}\rangle$ and $\vert e_{A},g_{B}\rangle$.

In the instantaneously inertial frame, the Hamiltonian describing the atom-field interaction is given by
\begin{equation}
H = \lambda \Big[m_{A}(\tau) \phi(x_{A}(\tau)) + m_{B}(\tau) \phi(x_{B}(\tau))\Big]\,,\label{ham}
\end{equation}
where $\lambda$ is the atom-field coupling constant assumed to be very small.
The forms of $\phi(x(\tau))$ and $m(\tau)$ are the same as given in eq.(s)(\ref{phiS}, \ref{mt}).
As a result of the atom-field interaction, transitions also occur for the two-atom system. According to the time-dependent perturbation theory in the first-order approximation, the transition amplitude for the atom-field system to transit from the initial state $\vert \chi, 0_{M}\rangle$ to the final state $\vert \chi', \phi_{f}\rangle$ is given by
\begin{equation}
\mathcal{A}_{\vert \chi, 0_{M}\rangle\rightarrow\vert \chi', \phi_{f}\rangle}=i\lambda \langle \chi',\phi_{f}\vert \int_{-\infty}^{+\infty}m_{A}(\tau) \phi(x_{A}(\tau))\vert \chi,0_{M}\rangle+A\rightleftharpoons B\,\text{term}\,.\label{transamp}
\end{equation}
Squaring the above transition amplitude and summing over all possible field states, the transition probability from the initial state $\vert \chi\rangle$ to the final state $\vert \chi'\rangle$ can be written as
\begin{equation}
\mathcal{P}_{\vert \chi\rangle\rightarrow \vert \chi' \rangle}=\lambda^2 \Big[\vert m^{(A)}_{\chi'\chi}\vert^2 F_{AA}(\Delta E)+m^{(B)}_{\chi'\chi}\,m^{(A)\,*}_{\chi'\chi}F_{AB}(\Delta E)\Big]+A\rightleftharpoons B\,\text{terms}\,,\label{transprob}
\end{equation}
where $m^{(A)}_{\chi'\chi}=\langle\chi'\vert m(0)\otimes \mathds{1}_{B}\vert\chi\rangle$, and  $m^{(B)}_{\chi'\chi}=\langle\chi'\vert \mathds{1}_{A}\otimes m(0)\vert\chi\rangle$.
The response function $F_{\xi\xi'}(\Delta E)$ is defined as 
\begin{equation}
F_{\xi\xi'}(\Delta E)=\int_{-\infty}^{+\infty} d\tau \int_{-\infty}^{+\infty} d\tau'\,e^{-i\Delta E(\tau-\tau')}\,G^{+}(x_{\xi}(\tau),x_{\xi'}(\tau'))\label{response}
\end{equation}
with $\xi,\xi'$ can be labeled by $A$ or $B$, and 
\begin{equation}
G^{+}(x_{\xi}(\tau),x_{\xi'}(\tau'))=\langle 0_{M}\vert\phi(x_{\xi}(\tau))\phi(x_{\xi'}(\tau'))\vert 0_{M}\rangle\label{wightman}
\end{equation}
is the Wightman function of the massless scalar field.
From equation \eqref{transprob} it is seen that for the two-atom system, the transition probability carries two terms one of them of which is associated with only one of the atoms and other one associated with both the atoms.

Exploiting the time translational invariance property of the Wightman function, the response function per unit proper time can be written as
\begin{equation}
\mathcal{F}_{\xi\xi'}(\Delta E)=\int_{-\infty}^{+\infty} d(\Delta\tau) \,e^{-i\Delta E\Delta\tau}\,G^{+}(x_{\xi}(\tau),x_{\xi'}(\tau'))\label{responserate}
\end{equation}
where $\Delta\tau=\tau-\tau'$.
Therefore, the  transition probability per unit proper time of the two-atom system from the initial state $\vert \chi\rangle$ to the final state $\vert \chi'\rangle$ turns out to be
\begin{equation}
\mathcal{R}_{\vert \chi\rangle\rightarrow \vert \chi' \rangle}=\lambda^2 \Big[\vert m^{(A)}_{\chi'\chi}\vert^2 \mathcal{F}_{AA}(\Delta E)+m^{(B)}_{\chi'\chi}\,m^{(A)\,*}_{\chi'\chi}\mathcal{F}_{AB}(\Delta E)\Big]+A\rightleftharpoons B\,\text{terms}\,.\label{transprobrate}
\end{equation}
The existence of the cross terms in the above equation  indicates that the rate of transition between the two neighbouring energy levels is not only related to the sum of the rates of transition of the two atoms, but also to their cross-correlation.

In the following section, this formalism is used to examine the rate of transitions of a two-atom system under various conditions of uniform acceleration and contact with the vacuum massless scalar field. The correlation between two atoms can be considered as a factor which can affect the transition rate of a two-atom system.  In practice, entanglement is not always maintained during experimental conditions, and non-maximally entangled states are frequently used as resources or probes. This motivates us to consider non-maximally entangled states, as well, in our analysis.
The  general pure quantum state of the two-atom system  is given by \cite{Chatterjee2021_1}
\begin{equation}
\vert \psi\rangle=\sin\theta\vert g_{A},e_{B}\rangle + \cos\theta \vert e_{A},g_{B}\rangle
\end{equation}
where the entanglement parameter $\theta$ lies in the range $0\leq\theta\leq\pi$. The above quantum state becomes a separable state for the values of $\theta=0,\pi/2,\pi$ and represents the maximally entangled state for the value $\theta=\pi/4$ (superradiant state), and $\theta=3\pi/4$ (subradiant state). Considering the above generic entangled quantum state as the initial state in the following section, we  explore how the non-inertial motion of the atoms, nature of the observer, type of the background field and the presence of the boundaries affect the rate of transition of the two-atom system. 
\section{Transition rates of the two-atom system from the viewpoint of a local inertial observer}\label{sec:Trate_in}
In this part, we analyse the transitions of a uniformly accelerated two-atom system prepared in any generic entangled state $\vert\psi\rangle$ that interacts with the vacuum massless scalar field from the perspective of a locally inertial observer. To see the boundary effects on the transitions of the uniformly accelerated two-atom system in this scenario, several cases have been studied in the following subsections.

\subsection{Transition rates for entangled atoms in empty space with respect to a local inertial observer}

Here we evaluate the transition rates of a two-atom system that is uniformly accelerating while interacting with a vacuum massless scalar field in the absence of any perfectly reflecting boundary. In the laboratory frame, trajectories of both the atoms read
\begin{equation}
t_{A}(\tau)=t_{B}(\tau)=\frac{1}{\alpha}\sinh(\alpha\tau),\,\,\,x_{A}(\tau)=x_{B}(\tau)=\frac{1}{\alpha}\cosh(\alpha\tau),\,\,\,y_{B}=y_{A}+d,\,\,\,z_{A}=z_{B}=0\,,\label{trajec}
\end{equation}
where $d$, $\alpha$ and $\tau$ denote the constant interatomic distance, proper acceleration and the proper time of the two-atom system.

Using the scalar field operator eq.\eqref{phiS} in eq.\eqref{wightman}, the Wightman function becomes \cite{Zhang2019}
\begin{align}
&G^{+}(x_{\xi}(\tau),x_{\xi'}(\tau'))\nonumber\\
=&-\frac{1}{4\pi^2}\frac{1}{(t_{\xi}(\tau)-t_{\xi'}(\tau')-i\varepsilon)^2-(x_{\xi}(\tau)-x_{\xi'}(\tau'))^2-(y_{\xi}(\tau)-y_{\xi'}(\tau'))^2-(z_{\xi}(\tau)-z_{\xi'}(\tau'))^2}\,.\label{wightmn1}
\end{align}
Substituting the trajectories (eq.\eqref{trajec}) in eq.\eqref{wightmn1}, the Wightman function turns out to be
\begin{equation}
 G^{+}(x_{\xi}(\tau),x_{\xi'}(\tau'))=-\frac{\alpha^2}{16\pi^2}\frac{1}{\sinh^2\left[\frac{1}{2}(\alpha\Delta\tau-i\varepsilon)\right]} \label{Wght1}
\end{equation}
with $\Delta\tau=\tau-\tau'$ for $\xi=\xi'$, and 
 \begin{equation}
 G^{+}(x_{\xi}(\tau),x_{\xi'}(\tau'))=-\frac{\alpha^2}{16\pi^2}\frac{1}{\sinh^2\left[\frac{1}{2}(\alpha\Delta\tau-i\varepsilon)\right]-\frac{1}{4}d^2\alpha^2} \label{Wght2}
 \end{equation}
 for $\xi\neq\xi'$.
Substituting the Wightman functions (eq.(s)(\ref{Wght1}, \ref{Wght2})) into eq.\eqref{response} and eq.\eqref{transprobrate}, the transition rate of the two-atom system
from the initial state $\vert \psi\rangle$ to the final state $\vert E_n\rangle$ can be expressed as
\begin{align}
\mathcal{R}_{\vert \psi\rangle\rightarrow \vert E_n\rangle}=\lambda^2 \Big[&\vert m^{(A)}_{E_n\psi}\vert^2 \mathcal{F}_{AA}(\Delta E)+\vert m^{(B)}_{E_n\psi}\vert^2 \mathcal{F}_{BB}(\Delta E)+m^{(B)}_{E_n\psi}\,m^{(A)\,*}_{E_n\psi}\mathcal{F}_{AB}(\Delta E)\nonumber\\
&+m^{(A)}_{E_n\psi}\,m^{(B)\,*}_{E_n\psi}\mathcal{F}_{BA}(\Delta E)\Big] \label{transprobrate1}
\end{align}
 with
\begin{equation}\label{Int_1}
\mathcal{F}_{\xi\xi'}(\Delta E)=
 -\frac{\alpha^2}{16\pi^2}\int_{-\infty}^{+\infty} d(\Delta\tau) \,e^{-i\Delta E\Delta\tau}\frac{1}{\sinh^2\left[\frac{1}{2}(\alpha\Delta\tau-i\varepsilon)\right]}
\end{equation}
for $\xi=\xi'$ and 
\begin{equation}\label{Int_2}
\mathcal{F}_{\xi\xi'}(\Delta E)= -\frac{\alpha^2}{16\pi^2}\int_{-\infty}^{+\infty} d(\Delta\tau) \,e^{-i\Delta E\Delta\tau}\frac{1}{\sinh^2\left[\frac{1}{2}(\alpha\Delta\tau-i\varepsilon)\right]-\frac{1}{4}d^2\alpha^2}
\end{equation} 
for $\xi\neq\xi'$.
We simplify the transition rate (eq.\eqref{transprobrate1}) by performing contour integration, leading to
 \begin{align}
\mathcal{R}_{\vert \psi\rangle\rightarrow \vert E_n\rangle}=&\lambda^2\left\{\theta\,(-\Delta E)\left(\frac{\vert \Delta E\vert}{2\pi}+\frac{\sin 2\theta \sin(\frac{2\vert \Delta E\vert}{\alpha}\sinh^{-1}(\frac{1}{2}\alpha d))}{2\pi d\sqrt{1+\frac{1}{4}d^2\alpha^2}}\right)\left(1+\frac{1}{e^{2\pi\vert \Delta E\vert/\alpha}-1}\right)\right.\nonumber\\
&+\left.\theta\,(\Delta E)\left(\frac{ \Delta E}{2\pi}+\frac{\sin 2\theta \sin(\frac{2\Delta E}{\alpha}\sinh^{-1}(\frac{1}{2}\alpha d))}{2\pi d\sqrt{1+\frac{1}{4}d^2\alpha^2}}\right)\left(\frac{1}{e^{2\pi\Delta E/\alpha}-1}\right)\right\} \label{transrateE}
\end{align}
where $\theta(\Delta E)$ is the Heaviside step function defined as
\begin{equation}
\theta(\Delta E)=
\begin{cases}
1,\,\,\,\Delta E>0,\\
0,\,\,\,\Delta E<0.
\end{cases}
\end{equation}
It follows that the two transition processes, namely, the upward and downward transition can take place for the two-atom system under uniform acceleration with the upward transition rate given by
\begin{equation}\label{up_emp}
\mathcal{R}_{\vert \psi\rangle\rightarrow \vert e_{A}e_{B}\rangle}=\lambda^2\left\{\left(\frac{\omega_{0}}{2\pi}+\frac{\sin 2\theta \sin(\frac{2\omega_{0}}{\alpha}\sinh^{-1}(\frac{1}{2}\alpha d))}{2\pi d\sqrt{1+\frac{1}{4}d^2\alpha^2}}\right)\left(\frac{1}{\exp{2\pi\omega_{0}/\alpha}-1}\right)\right\}\,,
\end{equation}
and the downward transition rate given by
\begin{equation}\label{dwn_emp}
\mathcal{R}_{\vert \psi\rangle\rightarrow \vert g_{A}g_{B}\rangle}=\lambda^2\left\{\left(\frac{\omega_{0}}{2\pi}+\frac{\sin 2\theta \sin(\frac{2\omega_{0}}{\alpha}\sinh^{-1}(\frac{1}{2}\alpha d))}{2\pi d\sqrt{1+\frac{1}{4}d^2\alpha^2}}\right)\left(1+\frac{1}{\exp{2\pi\omega_{0}/\alpha}-1}\right)\right\}\,.
\end{equation} 
Interestingly, the ratio of the upward and the downward transition rates for the two-atom case is identical with that of the single atom case (eq.\eqref{Ratio}).
\subsection{Transition rates for entangled atoms in a cavity with respect to a local inertial observer}

\begin{figure}[!htbp]
\centering
\includegraphics[scale=0.4]{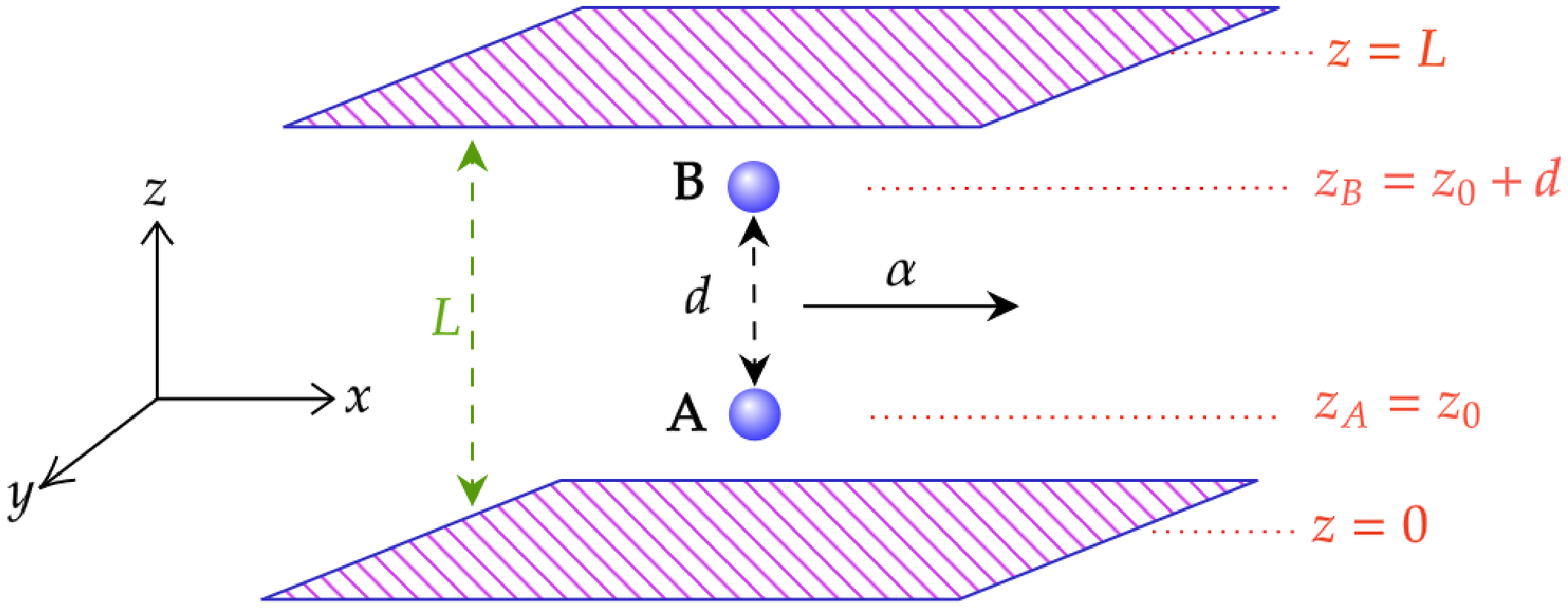}\label{fig:DoubleC_new}
\caption{Uniformly accelerated two-atom confined in a cavity.}
\end{figure}

Here we consider that a uniformly accelerated two-atom system interacts with a vacuum massless scalar field confined in a cavity having length $L$. Assuming that the two atoms are moving parallel to the boundary (see Fig. \ref{fig:DoubleC_new}) with their proper acceleration similar to the previous case,  we investigate the effect of the cavity on the transition rates.

Assuming the scalar field obeys the Dirichlet boundary condition $ \phi\vert_{z=0}=\phi\vert_{z=L}=0$, and using the method of images, the Wightman function of the massless scalar field confined to the cavity of length $L$ takes the form
\begin{align}
&G^{+}(x_{\xi}(\tau),x_{\xi'}(\tau'))\nonumber\\
=&-\frac{1}{4\pi^2}\displaystyle\sum_{n=-\infty}^{\infty}\left[\frac{1}{(t_{\xi}(\tau)-t_{\xi'}(\tau')-i\varepsilon)^2-(x_{\xi}(\tau)-x_{\xi'}(\tau'))^2-(y_{\xi}(\tau)-y_{\xi'}(\tau'))^2-(z_{\xi}(\tau)-z_{\xi'}(\tau')-nL)^2}\right.\nonumber\\
&\left.-\frac{1}{(t_{\xi}(\tau)-t_{\xi'}(\tau')-i\varepsilon)^2-(x_{\xi}(\tau)-x_{\xi'}(\tau'))^2-(y_{\xi}(\tau)-y_{\xi'}(\tau'))^2-(z_{\xi}(\tau)+z_{\xi'}(\tau')-nL)^2}\right] \label{WghtC}
\end{align}
To represent the atomic trajectories in terms of the atomic proper time $\tau$, we choose the Cartesian coordinates in the laboratory frame so that boundaries are fixed at $z = 0$ and $z=L$.

 Considering that the inter-atomic distance $d$ remains perpendicular while two atoms are moving parallel to the boundary with their proper acceleration, the atomic trajectories are given by
\begin{equation}
t_{A/B}(\tau)=\frac{1}{\alpha}\sinh(\alpha\tau),\,\,\,x_{A/B}(\tau)=\frac{1}{\alpha}\cosh(\alpha\tau),\,\,\,y_{A/B}=y_{0},\,\,z_{A}=z_{0},\,\,z_{B}=z_{0}+d\,.\label{trajecC}
\end{equation}
Using the above trajectories in eq.\eqref{WghtC}, the Wightman function becomes
\begin{equation}
 G^{+}(x_{\xi}(\tau),x_{\xi'}(\tau'))=-\frac{\alpha^2}{16\pi^2}\displaystyle\sum_{n=-\infty}^{\infty}\left[\frac{1}{\sinh^2\left[\frac{1}{2}(\alpha\Delta\tau-i\varepsilon)\right]-\frac{1}{4}d_{1}^2\alpha^2}-\frac{1}{\sinh^2\left[\frac{1}{2}(\alpha\Delta\tau-i\varepsilon)\right]-\frac{1}{4}d_{2}^2\alpha^2}\right] \label{WghtC1}
\end{equation}
for $\xi=\xi'$, with $d_1=nL,\,d_2=2z_{\xi}-nL$ and 
\begin{equation}
 G^{+}(x_{\xi}(\tau),x_{\xi'}(\tau'))=-\frac{\alpha^2}{16\pi^2}\displaystyle\sum_{n=-\infty}^{\infty}\left[\frac{1}{\sinh^2\left[\frac{1}{2}(\alpha\Delta\tau-i\varepsilon)\right]-\frac{1}{4}d_{3}^2\alpha^2}-\frac{1}{\sinh^2\left[\frac{1}{2}(\alpha\Delta\tau-i\varepsilon)\right]-\frac{1}{4}d_{4}^2\alpha^2}\right] \label{WghtC2}
\end{equation}
 for $\xi\neq\xi'$, with $d_3=-d-nL$ (for $\xi=A,\xi'=B$); $d_3=d-nL$ (for $\xi=B,\xi'=A$) and $d_4=2z_{0}+d-nL$.
 
Using the above Wightman functions, the rate of transition from the initial state $\vert \psi\rangle$ to the final state $\vert E_n\rangle$ can be written as
\begin{align}
\mathcal{R}_{\vert \psi\rangle\rightarrow \vert E_n\rangle}=\lambda^2 \displaystyle\sum_{n=-\infty}^{\infty}\Big[&\vert m^{(A)}_{E_n\psi}\vert^2 \mathcal{F}_{AA}(\Delta E)+\vert m^{(B)}_{E_n\psi}\vert^2 \mathcal{F}_{BB}(\Delta E)+m^{(B)}_{E_n\psi}\,m^{(A)\,*}_{E_n\psi}\mathcal{F}_{AB}(\Delta E)\nonumber\\
&+m^{(A)}_{E_n\psi}\,m^{(B)\,*}_{E_n\psi}\mathcal{F}_{BA}(\Delta E)\Big] \label{transprobrateC}
\end{align}
 with
\begin{equation}
\mathcal{F}_{\xi\xi'}(\Delta E)=
 -\frac{\alpha^2}{16\pi^2}\int_{-\infty}^{+\infty} d(\Delta\tau) \,e^{-i\Delta E\Delta\tau}\left[\frac{1}{\sinh^2\left[\frac{1}{2}(\alpha\Delta\tau-i\varepsilon)\right]-\frac{1}{4}d_{1}^2\alpha^2}-\frac{1}{\sinh^2\left[\frac{1}{2}(\alpha\Delta\tau-i\varepsilon)\right]-\frac{1}{4}d_{2}^2\alpha^2}\right]
\end{equation}
for $\xi=\xi'$ and 
\begin{equation}
\mathcal{F}_{\xi\xi'}(\Delta E)= -\frac{\alpha^2}{16\pi^2}\int_{-\infty}^{+\infty} d(\Delta\tau) \,e^{-i\Delta E\Delta\tau}\left[\frac{1}{\sinh^2\left[\frac{1}{2}(\alpha\Delta\tau-i\varepsilon)\right]-\frac{1}{4}d_{3}^2\alpha^2}-\frac{1}{\sinh^2\left[\frac{1}{2}(\alpha\Delta\tau-i\varepsilon)\right]-\frac{1}{4}d_{4}^2\alpha^2}\right]
\end{equation} 
for $\xi\neq\xi'$.
Eq.\eqref{transprobrateC} can be further simplified by performing contour integration  to obtain
\begin{align}
\mathcal{R}_{\vert \psi\rangle\rightarrow \vert E_n\rangle}=&\lambda^2\left\{\theta\,(-\Delta E)\left(\frac{\vert \Delta E\vert}{2\pi}+\mathfrak{f}\left(\vert \Delta E\vert,\,\alpha,\,\frac{L}{2}\right)-\cos^2\theta\,\mathfrak{h}\left(\vert \Delta E\vert,\,\alpha,\,z_{0},\,\frac{L}{2}\right)-\sin^2\theta\,\right.\right.\nonumber\\
&\times\left.\mathfrak{m}\left(\vert \Delta E\vert,\,\alpha,\,z_{0},\,d,\,\frac{L}{2}\right)+\sin2\theta\,\mathfrak{n}\left(\vert \Delta E\vert,\,\alpha,\,\frac{d}{2},\,\frac{L}{2}\right)-\sin2\theta\right.\nonumber\\
&\times \left.\mathfrak{m}\left(\vert \Delta E\vert,\,\alpha,\,z_{0},\,\frac{d}{2},\,\frac{L}{2}\right)\right)\left(1+\frac{1}{\exp{2\pi\vert \Delta E\vert/\alpha}-1}\right)\nonumber\\
&+\theta\,(\Delta E)\left(\frac{\Delta E}{2\pi}+\mathfrak{f}\left(\Delta E,\,\alpha,\,\frac{L}{2}\right)-\cos^2\theta\,\mathfrak{h}\left(\Delta E,\,\alpha,\,z_{0},\,\frac{L}{2}\right)-\sin^2\theta\,\right.\nonumber\\
&\times\left.\mathfrak{m}\left(\Delta E,\,\alpha,\,z_{0},\,d,\,\frac{L}{2}\right)+\sin2\theta\,\mathfrak{n}\left(\Delta E,\,\alpha,\,\frac{d}{2},\,\frac{L}{2}\right)-\sin2\theta\right.\nonumber\\
&\times \left.\left.\mathfrak{m}\left(\Delta E,\,\alpha,\,z_{0},\,\frac{d}{2},\,\frac{L}{2}\right)\right)\left(\frac{1}{\exp{2\pi\Delta E/\alpha}-1}\right)\right\} \label{transrateCE}
\end{align}
where we have defined
\begin{align}
\mathfrak{f}\left(\Delta E,\,\alpha,\,\frac{L}{2}\right)&=2\displaystyle\sum_{n=1}^{\infty}\mathfrak{g}\left(\Delta E,\,\alpha,\,\frac{nL}{2}\right)\,\\
\mathfrak{h}\left(\Delta E,\,\alpha,\,z_0\,,\frac{L}{2}\right)&=\displaystyle\sum_{n=-\infty}^{\infty}\mathfrak{g}\left(\Delta E,\,\alpha,\,z_0-\frac{nL}{2}\right)\,\\
\mathfrak{m}\left(\Delta E,\,\alpha,\,z_0\,,d,\,\frac{L}{2}\right)&=\displaystyle\sum_{n=-\infty}^{\infty}\mathfrak{g}\left(\Delta E,\,\alpha,\,z_0+d-\frac{nL}{2}\right)\,\\
\mathfrak{n}\left(\Delta E,\,\alpha,\,\frac{d}{2},\,\frac{L}{2}\right)&=\displaystyle\sum_{n=-\infty}^{\infty}\mathfrak{g}\left(\Delta E,\,\alpha,\,\frac{d-nL}{2}\right)\,\label{nfunc}
\end{align}
and $\mathfrak{g}\left(\Delta E,\,\alpha,\,z_0\right)$ is defined as
\begin{equation}
\mathfrak{g}\left(\Delta E,\,\alpha,\,z_0\right)=\frac{\sin(\frac{2\Delta E}{\alpha}\sinh^{-1}(\alpha z_0))}{4\pi z_0\sqrt{1+\alpha^2 z_0^2}}\,.\label{gfunc}
\end{equation}
From the above equation it follows that the two transition processes can take place for the two-atom system in the presence of a reflecting boundary with the upward transition rate given by
\begin{align}\label{up_C}
\mathcal{R}_{\vert \psi\rangle\rightarrow \vert e_{A}e_{B}\rangle}=&\lambda^2\left\{\left(\frac{\omega_0}{2\pi}+\mathfrak{f}\left(\omega_{0},\,\alpha,\,\frac{L}{2}\right)-\cos^2\theta\,\mathfrak{h}\left(\omega_{0},\,\alpha,\,z_{0},\frac{L}{2}\right)-\sin^2\theta\,\right.\right.\nonumber\\
&\times\left.\mathfrak{m}(\omega_{0},\,\alpha,\,z_{0},\,d,\,\frac{L}{2})+\sin2\theta\,\mathfrak{n}\left(\omega_{0},\,\alpha,\,\frac{d}{2},\,\frac{L}{2}\right)- \sin2\theta\right.\nonumber\\
&\times \left.\left.\mathfrak{m}\left(\omega_{0},\,\alpha,\,z_{0},\,\frac{d}{2},\,\frac{L}{2}\right)\right)\left(\frac{1}{\exp{2\pi\omega_{0}/\alpha}-1}\right)\right\}
\end{align}
and the downward transition rate given by
\begin{align}\label{dwn_C}
\mathcal{R}_{\vert \psi\rangle\rightarrow \vert g_{A}g_{B}\rangle}=&\lambda^2\left\{\left(\frac{\omega_0}{2\pi}+\mathfrak{f}\left(\omega_{0},\,\alpha,\,\frac{L}{2}\right)-\cos^2\theta\,\mathfrak{h}\left(\omega_{0},\,\alpha,\,z_{0},\frac{L}{2}\right)-\sin^2\theta\,\right.\right.\nonumber\\
&\times\left.\mathfrak{m}(\omega_{0},\,\alpha,\,z_{0},\,d,\,\frac{L}{2})+\sin2\theta\,\mathfrak{n}\left(\omega_{0},\,\alpha,\,\frac{d}{2},\,\frac{L}{2}\right)- \sin2\theta\right.\nonumber\\
&\times \left.\left.\mathfrak{m}\left(\omega_{0},\,\alpha,\,z_{0},\,\frac{d}{2},\,\frac{L}{2}\right)\right)\left(1+\frac{1}{\exp{2\pi\omega_{0}/\alpha}-1}\right)\right\}\,.
\end{align}
In order to obtain the single mirror and free space scenarios, we now take the limiting cases of these expressions. Taking the limit $L\rightarrow\infty$, we find that eq.(s)(\ref{up_C}, \ref{dwn_C}) reduce to the expression for the upward and the downward transition rate in the presence of a single reflecting boundary
\begin{align}
\mathcal{R}_{\vert \psi\rangle\rightarrow \vert e_{A}e_{B}\rangle}=&\lambda^2\Bigg\{\left(\frac{\omega_{0}}{2\pi}-\cos^2\theta\,\mathfrak{g}(\omega_{0},\,\alpha,\,z_{0})-\sin^2\theta\,\mathfrak{g}(\omega_{0},\,\alpha,\,(z_{0}+d))\right.\Bigg.\nonumber\\
&+\left.\left.\sin2\theta\left(\mathfrak{g}\left(\omega_{0},\,\alpha,\,\frac{d}{2}\right)-\mathfrak{g}\left(\omega_{0},\,\alpha,\,z_{0}+\frac{d}{2}\right)\right)\right)\left(\frac{1}{\exp{2\pi\omega_{0}/\alpha}-1}\right)\right\}
\end{align}
\begin{align}
\mathcal{R}_{\vert \psi\rangle\rightarrow \vert g_{A}g_{B}\rangle}=&\lambda^2\Bigg\{\left(\frac{\omega_{0}}{2\pi}-\cos^2\theta\,\mathfrak{g}(\omega_{0},\,\alpha,\,z_{0})-\sin^2\theta\,\mathfrak{g}(\omega_{0},\,\alpha,\,(z_{0}+d))\right.\Bigg.\nonumber\\
&+\left.\left.\sin2\theta\left(\mathfrak{g}\left(\omega_{0},\,\alpha,\,\frac{d}{2}\right)-\mathfrak{g}\left(\omega_{0},\,\alpha,\,z_{0}+\frac{d}{2}\right)\right)\right)\left(1+\frac{1}{\exp{2\pi\omega_{0}/\alpha}-1}\right)\right\}
\end{align}
It may be noted that the above relations resemble those of the single boundary results given in \cite{Zhang2019} but they are not identical as in our case the interatomic distance is perpendicular to the reflecting boundary whereas in \cite{Zhang2019} the interatomic distance is parallel to the reflecting boundary.
Similarly, taking the limits $L\rightarrow\infty$ and $z_{0}\rightarrow\infty$, eq.(s)(\ref{up_C}, \ref{dwn_C}) lead to the expressions for the upward and the downward transition rate in free space given by eq.(s)(\ref{up_emp}, \ref{dwn_emp}).

\begin{figure}[H]
\begin{minipage}{0.5\textwidth}
\centering
\includegraphics[scale=0.6]{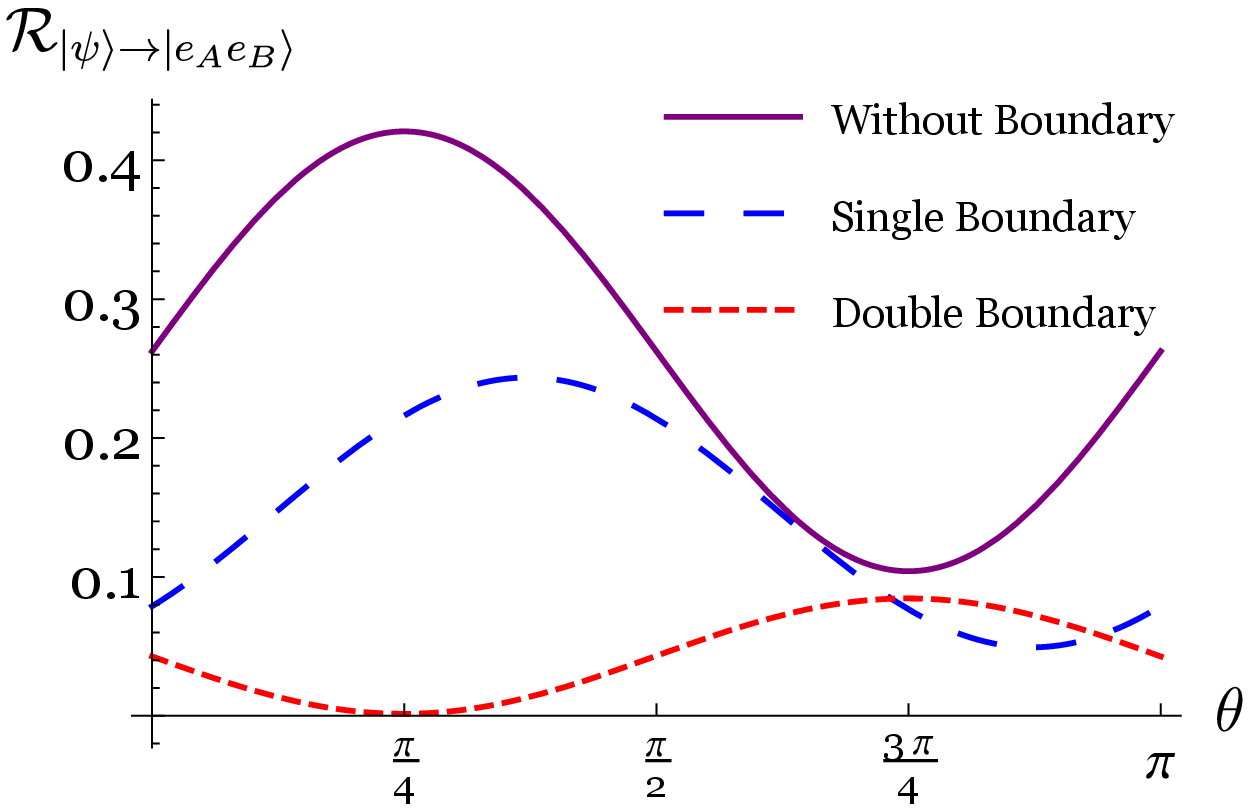}
\subcaption{Upward transition rate}\label{fig:PDouble_th}
\end{minipage}
\begin{minipage}{0.5\textwidth}
\centering
\includegraphics[scale=0.6]{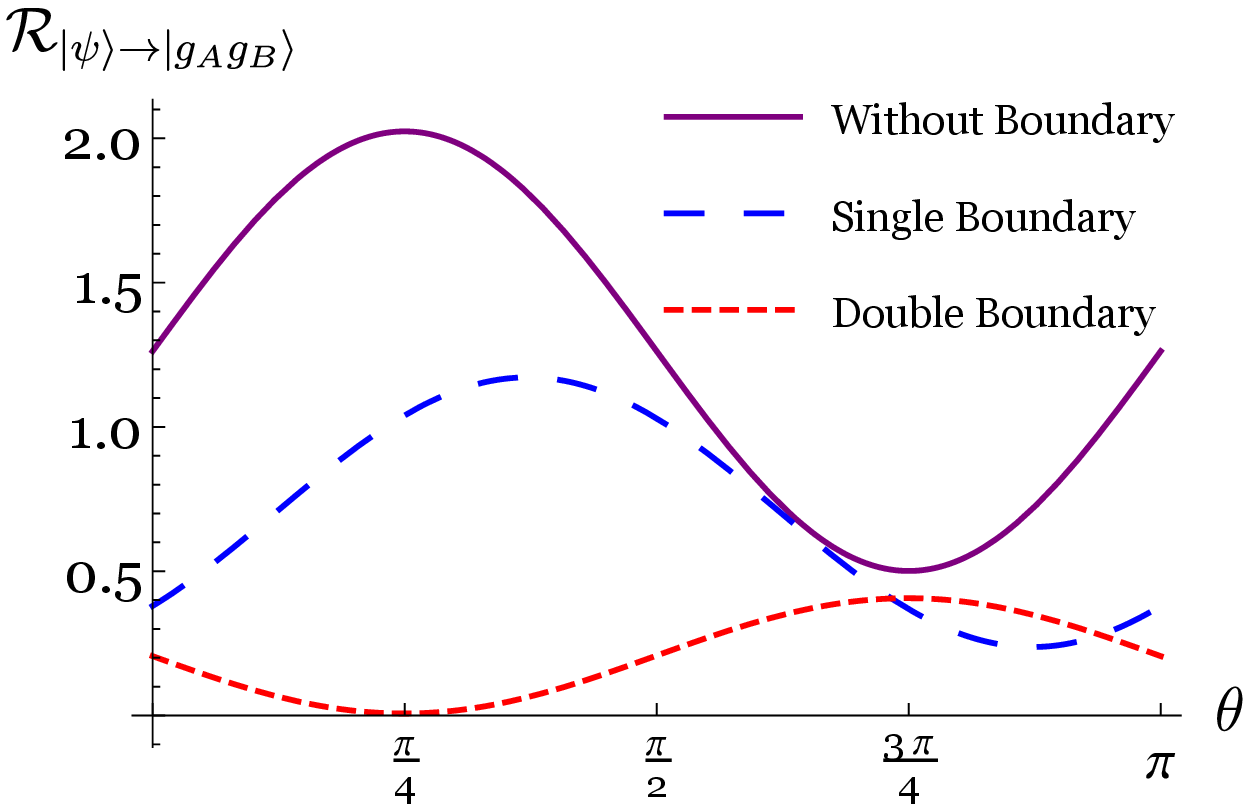}
\subcaption{Downward transition rate}\label{fig:PDouble_th1}
\end{minipage}
\caption{Transition rate (per unit $\frac{\lambda^2\omega_0}{2\pi}$) versus entanglement parameter for a fixed value of $\alpha/\omega_0=4,\,\omega_{0}d=0.5,\,\omega_{0}L=1.2,\,\omega_{0}z_0=0.2$.}
\end{figure}

We now study the variation of the transition rate of an entangled two atom system from an initial entangled state $\vert \psi\rangle$ to a product state with higher energy value $\vert e_{A}e_{B}\rangle$ confined to a cavity with the atomic acceleration ($\alpha$), length of the cavity ($L$), distance of any one atom from one boundary ($z_0$), entanglement parameter ($\theta$), and the interatomic distance ($d$). The findings are plotted below, where all physical quantities are expressed in dimensionless units. Since cavity effects are significant when the length scales are comparable \cite{Donaire}, hence we choose a similar order of magnitude for $\omega_{0}L,\,\omega_{0}z_{0}\,\text{and}\,\omega_{0}d$.  In Figure \ref{fig:PDouble_th}, we show the behaviour of the transition rate with respect to the entanglement parameter for the cases where the atoms are in free space, in the vicinity of a single boundary and inside a cavity.

 From the above Figure, it can be seen that the transition rate $\vert \psi\rangle\rightarrow\vert e_{A}e_{B}\rangle$ (per unit $\frac{\lambda^2\omega_0}{2\pi}$) varies sinusoidally with the entanglement parameter $\theta$. In free space, the transition rate increases (from the case corresponding to the zero entanglement product state) with  increase in the entanglement parameter and it becomes maximum when the initial state is maximally entangled ($\theta=\pi/4$  super-radiant state). Further increment of the entanglement parameter decreases the transition rate and it becomes minimum at $\theta=3\pi/4$ (sub-radiant state). In the vicinity of a single boundary, behaviour of the transition rate is quite similar to the free space scenario, with a slight shifting of the extremum points). Surprisingly, inside the cavity, the behaviour of the transition rate is opposite to the free space scenario. The transition rate decreases with the increase in entanglement parameter and it vanishes at $\theta=\pi/4$. Thus, the $\theta=\pi/4$ state exhibits a subradiant behaviour for the cavity set up. Further increment of entanglement parameter increases the transition rate and it becomes maximum at $\theta=3\pi/4$. Note also, that around $\theta=3\pi/4$, the values of
the transition rate corresponding to cases of empty space, single boundary, and two boundaries, are nearly the same. 

Figure \ref{fig:PDouble_th1} for the downward transition rates  shows a similar behaviour  with respect to the entanglement parameter. The only difference is that the magnitude of the downward transition rate is greater than the upward transition rate. From both the Figures, it is observed that when the initial entangled state is the super-radiant maximally entangled state ($\theta=\pi/4$), then there is no upward or downward transition 
inside a
cavity. Therefore, this observation indicates that for this value of the entanglement parameter, the entanglement of the initial state is preserved.

\begin{figure}[H]
\centering
\includegraphics[scale=0.6]{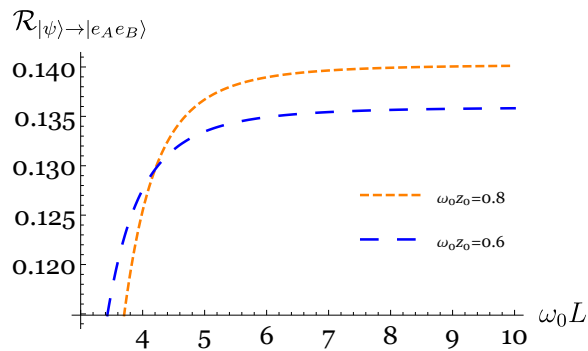}
\caption{Transition rate from $\vert \psi\rangle\rightarrow\vert e_{A}e_{B}\rangle$ (per unit $\frac{\lambda^2\omega_0}{2\pi}$) versus separation between two boundaries, $\alpha/\omega_0=4$.}\label{fig:PDouble_L}
\end{figure}

Figure \ref{fig:PDouble_L} shows the variation of the transition rate from $\vert \psi\rangle\rightarrow\vert e_{A}e_{B}\rangle$ (per unit $\frac{\lambda^2\omega_0}{2\pi}$) with respect to the length of the cavity for different values of distance of any one atom from one boundary. From the plots, it can be seen that for a fixed value of the initial atomic distance $z_0$ of any one atom from the nearest boundary, the transition rate get enhanced when the cavity length increases and attains a maximum value for large values of $L$ ($\omega_0 L>>\omega_0 z_0$). This behaviour is similar
to that of the single atom case, as  mentioned earlier. As more number of field modes take part in the interaction between the scalar field and the  atoms due to the increased cavity length, the transition rate increases. When $\omega_0 L>>\omega_0 z_0$,  the cavity scenario reduces to a single boundary set up and hence the upward transition rate takes a constant value. It is also observed that the saturation value of the transition rate is more for a larger value of $\omega_0 z_0$.

\begin{figure}[H]
\centering
\includegraphics[scale=0.6]{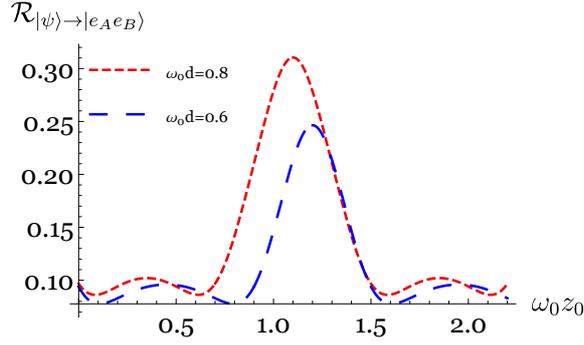}
\caption{Transition rate from $\vert \psi\rangle\rightarrow\vert e_{A}e_{B}\rangle$ (per unit $\frac{\lambda^2\omega_0}{2\pi}$) versus distance of any one atom from one boundary, $\alpha/\omega_0=4,\,\omega_{0}L=3$.}\label{fig:PDouble_z}
\end{figure}

Figure \ref{fig:PDouble_z} shows the variation of the transition rate from $\vert \psi\rangle\rightarrow\vert e_{A}e_{B}\rangle$ (per unit $\frac{\lambda^2\omega_0}{2\pi}$) with respect to the distance of any one atom from one boundary $\omega_0 z_0$ for different interatomic distances 
$\omega_0 d$. From the plots, it is observed that for a fixed value of the interatomic distance and cavity length, the transition rate is much smaller when the atoms are close to the any one of the boundaries. Thereafter, with increase of the atomic distance from one boundary, the transition rate 
rises sharply and reaches maximum when the distance of both atoms to their nearest boundaries are equal. The importance of boundary effects are thereby
clearly exhibited. From the plot, it is also seen that increasing the interatomic distance increases the upward transition rate for a fixed cavity length.

\begin{figure}[H]
\centering
\includegraphics[scale=0.6]{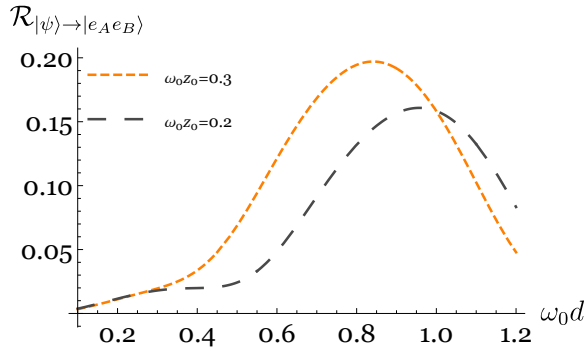}
\caption{Transition rate from $\vert \psi\rangle\rightarrow\vert e_{A}e_{B}\rangle$ (per unit $\frac{\lambda^2\omega_0}{2\pi}$) versus interatomic distance, $\alpha/\omega_0=4,\,\omega_{0}L=1.5$.}\label{fig:PDouble_d}
\end{figure}

Figure \ref{fig:PDouble_d} shows the variation of the transition rate from $\vert \psi\rangle\rightarrow\vert e_{A}e_{B}\rangle$ (per unit $\frac{\lambda^2\omega_0}{2\pi}$) with respect to the interatomic distance for different values of distance of any one atom from one boundary. From the plots, we see that for a fixed atomic distance from one boundary and cavity length, transition rate initially increases when the interatomic distance increases. After a certain value of interatomic distance, increasing the distance between the two atoms further make them move closer to the boundary and hence due to boundary effects, the transition rate falls down.

\begin{figure}[H]
\centering
\includegraphics[scale=0.6]{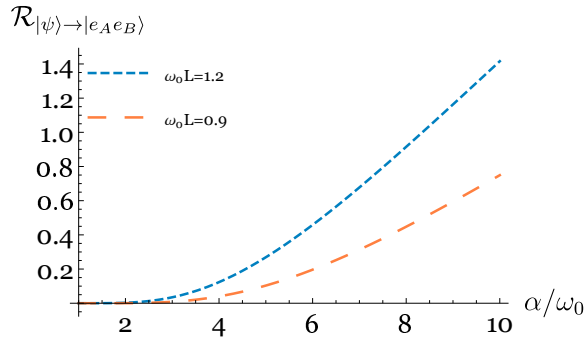}
\caption{Transition rate from $\vert \psi\rangle\rightarrow\vert e_{A}e_{B}\rangle$ (per unit $\frac{\lambda^2\omega_0}{2\pi}$) versus acceleration, $\omega_{0}d=0.5,\,\omega_{0}z_0=0.3$.}\label{fig:PDouble_a}
\end{figure}

Figure \ref{fig:PDouble_a} shows the variation of the transition rate from $\vert \psi\rangle\rightarrow\vert e_{A}e_{B}\rangle$ (per unit $\frac{\lambda^2\omega_0}{2\pi}$) with respect to the atomic acceleration for different values of cavity length. Similar to the single atom case, it is seen that when the atomic acceleration is  increased, the transition rate also increases and the rate of transition depends on the cavity length. This is expected since acceleration radiation should increase with increase in acceleration.

\begin{figure}[H]
\begin{minipage}{0.5\textwidth}
\centering
\includegraphics[scale=0.6]{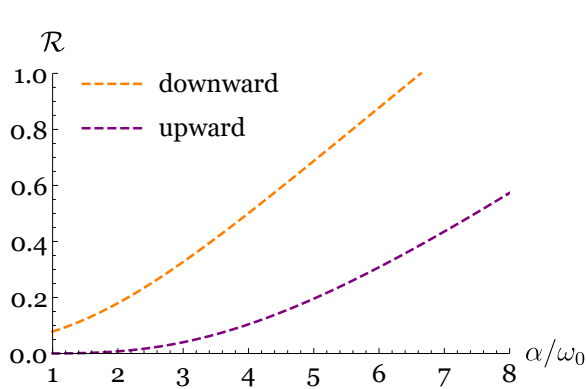}
\subcaption{In free space}\label{fig:R_free}
\end{minipage}
\begin{minipage}{0.5\textwidth}
\centering
\includegraphics[scale=0.6]{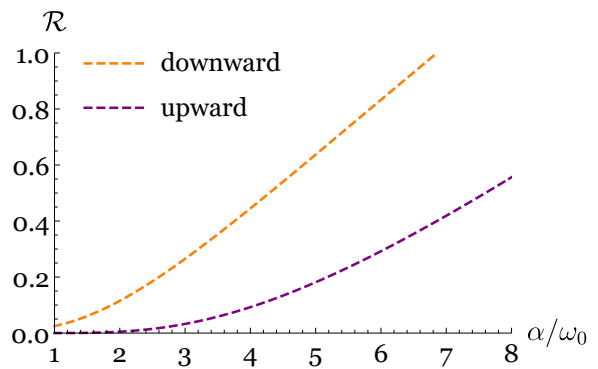}
\subcaption{Inside the cavity for a fixed value of $\omega_{0}L=4,\,\omega_{0}z_0=1$}\label{fig:R_cavity}
\end{minipage}
\caption{Transition rate (per unit $\frac{\lambda^2\omega_0}{2\pi}$) versus acceleration for a fixed value of $\theta=3\pi/4,\,\omega_{0}d=0.5$.}\label{fig:R_comp}
\end{figure}

In Figure \ref{fig:R_comp}, we have plotted the upward and the downward transition rates with respect to the atomic acceleration for two cases, namely, two atoms are in free space (Figure \ref{fig:R_free}), two atoms are confined to a cavity (Figure \ref{fig:R_cavity}). A comparison of the two plots reveals that  the  downward transition rate can get diminished for
a suitable choice of parameters when the atoms are inside the cavity. The
upward transitions in both cases are driven by the acceleration, as is clear from the corresponding expresions, as well.

Before going to the next section, we present a quantitative estimation of the upward transition rate for the two-atom system, composed of two Rubidium atoms $Rb^{87}$ placed inside a cavity. Following \cite{Chatterjee2021_1}, we choose the length of the cavity in the order of $100nm$, distance between any one atom and the nearest boundary in the order of $20nm$, interatomic distance in the order of $30nm$, energy gap between the most generic entangled state and excited state of the two-atom system is of the order of $0.5eV$, and the acceleration in the order of $10^{17}m/s^2$. Using eq.\eqref{up_C} with the coupling constant $\lambda=0.1$, taking the entanglement parameter $\theta=3\pi/4$ for the maximally entangled state, and the above values, the upward transition rate of the uniformly accelerated two-atom system inside a cavity becomes $3.75\times 10^{-12}eV=5.68\times10^{3}s^{-1}$.

\section{Transition rates of the two-atom system from the viewpoint of a coaccelerated observer}\label{sec:Trate_co}
In this section, the transitions of a uniformly accelerated two-atom system prepared in any generic entangled state $\vert\psi\rangle$ that interacts with a massless scalar field is analysed from the perspective of a coaccelerated observer. To see the boundary effects on the transitions of the uniformly accelerated two-atom system in this scenario, we consider that the coordinate of the coaccelerated frame will be the Rindler coordinate $(\tau,\,\eta,\,y,\,z)$ with the relation with those of the laboratory coordinates ($t,\,x,\,y,\,z $) being given by
\begin{equation}
t(\tau,\eta)=\frac{1}{\alpha}e^{\alpha\eta}\sinh(\alpha\tau),\,\,\,x(\tau,\eta)=\frac{1}{\alpha}e^{\alpha\eta}\cosh(\alpha\tau)\,.
\end{equation}
In the coaccelerated frame, the field operator $\phi(x(\tau))$ is replaced by its Rindler counterpart $\bar{\phi}(x(\tau))$ and  takes the form
\begin{equation}
\bar{\phi}(\tau,\mathbf{x}))=\int_{0}^{\infty}d\omega\int_{-\infty}^{\infty}dk_{y}\int_{-\infty}^{\infty}dk_{z}\left[b_{\omega,k_{y},k_{z}}\mathcal{V}_{\omega,k_{y},k_{z}}(\tau,\mathbf{x})+b^{\dagger}_{\omega,k_{y},k_{z}}\mathcal{V}^{\star}_{\omega,k_{y},k_{z}}(\tau,\mathbf{x})\right]\label{phi_co}
\end{equation}
with
\begin{equation}
\mathcal{V}_{\omega,k_{y},k_{z}}(\tau,\mathbf{x})=\sqrt{\frac{\sinh(\pi\omega/\alpha)}{4\pi^{4}\alpha}}\mathcal{K}_{i\frac{\omega}{\alpha}}\left(\frac{k_{\perp}}{\alpha}e^{\alpha\eta}\right)e^{-i\omega\tau+ik_{y}y+ik_{z}z}\label{modefn}
\end{equation}
being the positive frequency orthonormal mode solution, $\mathcal{K}_{\nu}(x)$ is the Bessel function of imaginary argument and $k_{\perp}\equiv\vert\mathbf{k}_{\perp}\vert=\sqrt{k_{y}^2+k_{z}^2}$.
The interaction between the atoms and the scalar field is given by \cite{Zhou2020}
\begin{equation}
H = \lambda \Big[m_{A}(\tau) \bar{\phi}(x_{A}(\tau)) + m_{B}(\tau) \bar{\phi}(x_{B}(\tau))\Big]\,.\label{ham1}
\end{equation}
To determine the transition rate of the two-atom system in the coaccelerated frame, we a thermal field  at an arbitrary temperature $T$. As the thermal state is a mixed state, in order to calculate the response of the two atoms coupled to the massless scalar field, additionally it is assumed that the field state can be represented by a pure state $\vert\sigma_{\omega,k_{y},k_{z}}\rangle$ with a probability factor $p_{\sigma}(\omega)=e^{-\beta\omega\sigma}/N(\omega)$ with $\beta=1/T$ and $N(\omega)=\displaystyle\sum_{\sigma=0}^{\infty}e^{-\beta\omega\sigma}$. In this case, $\vert \psi,\,\sigma_{\omega,k_{y},k_{z}}\rangle$ and $\vert E_n,\,\gamma_{\omega',k^{\prime}_{y},k^{\prime}_{z}}\rangle$ can be used to represent the initial and the final state of the atom-field system.

Following the procedure described in the previous sections, the probability that the atom-field system will transit from initial state $\vert \psi\rangle$ to final state $\vert E_n\rangle$ is then given by 
\begin{equation}
\mathcal{P}_{\vert \psi_{\pm}\rangle\rightarrow\vert E_n\rangle}=\lambda^2 \Big[\vert m^{(A)}_{E_n\psi_{\pm}}\vert^2 F^{\beta}_{AA}(\Delta E)+m^{(B)}_{E_n\psi_{\pm}}\,m^{(A)\,*}_{E_n\psi_{\pm}}F^{\beta}_{AB}(\Delta E)\Big]+A\rightleftharpoons B\,\text{terms}\,.
\end{equation}
The response function $F^{\beta}_{\xi\xi'}(\Delta E)$ is defined as 
\begin{equation}
F^{\beta}_{\xi\xi'}(\Delta E)=\int_{-\infty}^{+\infty} d\tau \int_{-\infty}^{+\infty} d\tau'\,e^{-i\Delta E(\tau-\tau')}\,G^{+}_{\beta}(x_{\xi}(\tau),x_{\xi'}(\tau'))\label{response-co}
\end{equation}
with $\xi,\xi'$  labeled by $A$ or $B$, and 
\begin{align}
&G^{+}_{\beta}(x_{\xi}(\tau),x_{\xi'}(\tau'))=\frac{tr[\rho'\phi(x_{\xi}(\tau))\phi(x_{\xi'}(\tau')]}{tr[\rho']}\nonumber\\
&=N^{-1}(\omega)\displaystyle\sum_{\sigma=0}^{\infty}\int_{0}^{\infty}d\omega\int_{-\infty}^{\infty}dk_{y}\int_{-\infty}^{\infty}dk_{z}e^{-\beta\omega\sigma}\langle \sigma_{\omega,k_{y},k_{z}}\vert\bar{\phi}(x_{\xi}(\tau))\bar{\phi}(x_{\xi'}(\tau'))\vert \sigma_{\omega,k_{y},k_{z}}\rangle\label{wightman-co}
\end{align}
is the positive frequency Wightman function of the scalar field in a thermal state at an arbitrary temperature $T$ in the coaccelerated frame.
Exploiting the time translational invariance property of the Wightman function, the response function per unit proper time can be written as
\begin{equation}
\mathcal{F}^{\beta}_{\xi\xi'}(\Delta E)=\int_{-\infty}^{+\infty} d(\Delta\tau) \,e^{-i\Delta E\Delta\tau}\,G^{+}_{\beta}(x_{\xi}(\tau),x_{\xi'}(\tau'))\,.\label{responserate-co}
\end{equation}
Therefore, the transition probability per unit proper time of the two-atom system from the initial state $\vert \chi\rangle$ to the final state $\vert \chi'\rangle$ turns out to be
\begin{equation}
\mathcal{R}^{\beta}_{\vert \chi\rangle\rightarrow \vert \chi' \rangle}=\lambda^2 \Big[\vert m^{(A)}_{\chi'\chi}\vert^2 \mathcal{F}^{\beta}_{AA}(\Delta E)+m^{(B)}_{\chi'\chi}\,m^{(A)\,*}_{\chi'\chi}\mathcal{F}^{\beta}_{AB}(\Delta E)\Big]+A\rightleftharpoons B\,\text{terms}\,.\label{transprobrate-co}
\end{equation}
\subsection{Transition rates for entangled atoms in empty space with respect to a coaccelerated observer}
Substituting eq.\eqref{phi_co} into eq.\eqref{wightman-co}, the thermal Wightman function takes the following form for an arbitrary temperature $T$:
\begin{align}
&G^{+}_{\beta}(x_{\xi}(\tau),x_{\xi'}(\tau'))\nonumber\\
&=\int_{0}^{\infty}d\omega\int_{-\infty}^{\infty}dk_{y}\int_{-\infty}^{\infty}dk_{z}\bigg[\displaystyle\sum_{\sigma=0}^{\infty}(\sigma+1)e^{-\beta\omega\sigma}\mathcal{V}_{\omega,k_{y},k_{z}}(\tau_{\xi},\mathbf{x}_{\xi})\mathcal{V}^{\star}_{\omega,k_{y},k_{z}}(\tau^{\prime}_{\xi'},\mathbf{x}^{\prime}_{\xi'})\bigg.\nonumber\\
&+\bigg.\displaystyle\sum_{\sigma=1}^{\infty}\sigma e^{-\beta\omega\sigma}\mathcal{V}^{\star}_{\omega,k_{y},k_{z}}(\tau_{\xi},\mathbf{x}_{\xi})\mathcal{V}_{\omega,k_{y},k_{z}}(\tau^{\prime}_{\xi},\mathbf{x}^{\prime}_{\xi'})\bigg]\bigg/\displaystyle\sum_{\sigma=0}^{\infty}e^{-\beta\omega\sigma}\nonumber\\
&=\int_{0}^{\infty}d\omega\int_{-\infty}^{\infty}dk_{y}\int_{-\infty}^{\infty}dk_{z}\bigg[\frac{e^{\omega/T}}{e^{\omega/T}-1}\mathcal{V}_{\omega,k_{y},k_{z}}(\tau_{\xi},\mathbf{x}_{\xi})\mathcal{V}^{\star}_{\omega,k_{y},k_{z}}(\tau^{\prime}_{\xi'},\mathbf{x}^{\prime}_{\xi'})\bigg.\nonumber\\
&+\bigg.\frac{1}{e^{\omega/T}-1}\mathcal{V}^{\star}_{\omega,k_{y},k_{z}}(\tau_{\xi},\mathbf{x}_{\xi})\mathcal{V}_{\omega,k_{y},k_{z}}(\tau^{\prime}_{\xi},\mathbf{x}^{\prime}_{\xi'})\bigg]\label{wightman-co1}
\end{align}
In the Rindler coordinates, the trajectories of both the atoms are given by
\begin{equation}
t_{A/B}=\tau,\,\,\,\eta_{A/B}=0,\,\,\,y_{A/B}=0,\,\,z_{A}=0\,\,z_{B}=d\,.\label{trajec_co}
\end{equation}
Substituting eq.(s)(\ref{modefn}, \ref{trajec_co}) into eq.\eqref{wightman-co1}, the thermal Wightman function takes the form
\begin{align}
&G^{+}_{\beta}(x_{\xi}(\tau),x_{\xi'}(\tau'))\nonumber\\
&=\frac{1}{4\pi^{4}\alpha}\int_{0}^{\infty}d\omega\int_{-\infty}^{\infty}dk_{y}\int_{-\infty}^{\infty}dk_{z}\sinh\left(\frac{\pi\omega}{\alpha}\right)\mathcal{K}_{i\omega/\alpha}^{2}\left(\frac{k_{\perp}}{\alpha}\right)\bigg[\frac{e^{\omega/T}}{e^{\omega/T}-1}e^{-i\omega(\tau-\tau')}\bigg.\nonumber\\
&+\bigg.\frac{1}{e^{\omega/T}-1}e^{i\omega(\tau-\tau')}\bigg]\,.\label{wightman-co2}
\end{align}
Now, using following integrals
\begin{equation}
\int_{-\infty}^{\infty}dk_{y}\int_{-\infty}^{\infty}dk_{z}\mathcal{K}_{i\omega/\alpha}^{2}\left(\frac{k_{\perp}}{\alpha}\right)=\frac{\alpha\pi^2\omega}{\sinh(\pi\omega/\alpha)}
\end{equation}
and
\begin{equation}
\int_{-\infty}^{\infty}dk_{y}\int_{-\infty}^{\infty}dk_{z}\mathcal{K}_{i\omega/\alpha}^{2}\left(\frac{k_{\perp}}{\alpha}\right)e^{-ik_{z}d}=\frac{\alpha\pi^2}{\sinh(\pi\omega/\alpha)}\frac{\sin\left(\frac{2\omega}{\alpha}\sinh^{-1}\left(\frac{d\alpha}{2}\right)\right)}{d\sqrt{1+\frac{1}{4}\alpha^2 d^2}}
\end{equation}
the thermal Wightman function  takes the form
\begin{equation}\label{wightman-cof}
G^{+}_{\beta}(x_{\xi}(\tau),x_{\xi'}(\tau'))
=-\frac{1}{4\pi^{2}}\displaystyle\sum_{s=-\infty}^{\infty}\frac{1}{(\Delta\tau-is\beta-i\varepsilon)^2}
\end{equation}
for $\xi=\xi'$,
and
\begin{equation}\label{wightman-cof1}
G^{+}_{\beta}(x_{\xi}(\tau),x_{\xi'}(\tau'))=-\frac{\mathcal{B}}{4\pi^{2}\mathcal{C}}\displaystyle\sum_{s=-\infty}^{\infty}\frac{1}{(\Delta\tau-is\beta-i\varepsilon)^2-\mathcal{B}^2}
\end{equation}
with $\mathcal{B}=\frac{2}{\alpha}\sinh^{-1}\left(\frac{d\alpha}{2}\right)$ and $\mathcal{C}=d\sqrt{1+\frac{1}{4}\alpha^2 d^2}$, for $\xi\neq\xi'$.
By inserting the aforementioned Wightman functions into eq.(s)(\ref{responserate-co}, \ref{transprobrate-co}), and performing the integrations using contour integration, the upward and downward transition rates of the two-atom system submerged in the thermal bath turn out to be
\begin{equation}\label{upco_emp}
\mathcal{R}^{\beta}_{\vert \psi\rangle\rightarrow \vert e_{A}e_{B}\rangle}=\lambda^2\left\{\left(\frac{\omega_{0}}{2\pi}+\frac{\sin 2\theta \sin(\frac{2\omega_{0}}{\alpha}\sinh^{-1}(\frac{1}{2}\alpha d))}{2\pi d\sqrt{1+\frac{1}{4}d^2\alpha^2}}\right)\left(\frac{1}{\exp{\omega_{0}/T}-1}\right)\right\}
\end{equation}
\begin{equation}\label{dwnco_emp}
\mathcal{R}^{\beta}_{\vert \psi\rangle\rightarrow \vert g_{A}g_{B}\rangle}=\lambda^2\left\{\left(\frac{\omega_{0}}{2\pi}+\frac{\sin 2\theta \sin(\frac{2\omega_{0}}{\alpha}\sinh^{-1}(\frac{1}{2}\alpha d))}{2\pi d\sqrt{1+\frac{1}{4}d^2\alpha^2}}\right)\left(1+\frac{1}{\exp{\omega_{0}/T}-1}\right)\right\}\,.
\end{equation}
From the above equations it follows that in the coaccelerated frame both the upward and the downward transitions can occur for the two-atom system immersed in the thermal bath which is very similar to the transitions observed by an instantaneously inertial observer. Taking the limiting value of the temperature of the coaacelerated frame $T\rightarrow 0$, here also we can see that the upward transition rate vanishes, which is consistent with that in the Minkowski vacuum (eq.(\ref{up_emp})). Eq.(s)(\ref{up_emp}, \ref{dwn_emp}, \ref{upco_emp}, \ref{dwnco_emp}) indicate that the transition rates of the uniformly accelerated two-atom system in the  generic entangled state seen by an instantaneously inertial observer and by a coaccelerated observer are identical only  if the thermal bath temperature in the coaccelerated frame is equal to the FDU temperature $T=\alpha/2\pi$.
\subsection{Transition rates for entangled atoms in a cavity with respect to a coaccelerated observer}
Here we consider that a uniformly accelerated two-atom system interacts with a massless scalar field confined to a cavity of length $L$ from the perspective of a coaccelerated observer. We assume that two perfectly reflecting boundaries are placed at $z=0$ and $z=L$. As in the case of  the single atom, 
the scenario here too depicts  a static two-atom system interacting with a massless scalar field in a thermal state at an arbitrary temperature $T$ inside a cavity of length $L$.

Considering that the scalar field obeys the Dirichlet boundary condition $\phi\vert_{z=0}=\phi\vert_{z=L}=0$, and by following a similar method we have used for the single atom case, the positive frequency thermal Wightmann function of the massless scalar field confined in the cavity turns out to be
\begin{align}\label{wightman-coC}
&G^{+}_{\beta}(x_{\xi}(\tau),x_{\xi'}(\tau'))\nonumber\\
&=\frac{1}{4\pi^{4}\alpha}\displaystyle\sum_{n=-\infty}^{\infty}\int_{0}^{\infty}d\omega\int_{-\infty}^{\infty}dk_{y}\int_{-\infty}^{\infty}dk_{z}\sinh\left(\frac{\pi\omega}{\alpha}\right)\mathcal{K}_{i\omega/\alpha}\left(\frac{k_{\perp}}{\alpha}e^{\alpha\eta_{\xi}}\right)\mathcal{K}_{i\omega/\alpha}\left(\frac{k_{\perp}}{\alpha}e^{\alpha\eta^{\prime}_{\xi'}}\right)\nonumber\\
&\times\bigg\{\cos\big[k_{z}(z_{\xi}-z^{\prime}_{\xi'}-nL)\big]-\cos\big[k_{z}(z_{\xi}+z^{\prime}_{\xi'}-nL)\big]\bigg\}\bigg[\frac{e^{\omega/T}}{e^{\omega/T}-1}e^{-i\omega(\tau_{\xi}-\tau^{\prime}_{\xi'})+ik_{y}(y_{\xi}-y^{\prime}_{\xi'})}\bigg.\nonumber\\
&+\bigg.\frac{1}{e^{\omega/T}-1}e^{i\omega(\tau_{\xi}-\tau^{\prime}_{\xi'})-ik_{y}(y_{\xi}-y^{\prime}_{\xi'})}\bigg]\,.
\end{align}

\begin{figure}[!htbp]
\centering
\includegraphics[scale=0.4]{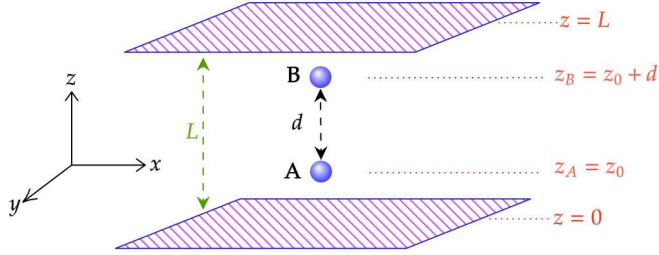}
\caption{Static two-atom confined in a cavity.}
\end{figure}\label{fig:DoubleC1_new}

Considering the inter-atomic distance $d$ to remain perpendicular while 
the two atoms are moving parallel to the boundaries with their proper acceleration, the atomic trajectories are given by
\begin{equation}
t_{A/B}=\tau,\,\,\,\eta_{A/B}=0,\,\,\,y_{A/B}=0,\,\,z_{A}=z_0\,\,z_{B}=z_{0}+d\,.\label{trajec_coC}
\end{equation}
Inserting the above trajectories in eq.\eqref{wightman-coC} and following the procedure mentioned in the previous subsection, the thermal Wightman function becomes
\begin{equation}
G^{+}_{\beta}(x_{\xi}(\tau),x_{\xi'}(\tau'))=-\frac{1}{4\pi^{2}}\displaystyle\sum_{n=-\infty}^{\infty}\displaystyle\sum_{s=-\infty}^{\infty}\left[\frac{\mathcal{B}_1}{\mathcal{C}_1}\frac{1}{(\Delta\tau-is\beta-i\varepsilon)^2-\mathcal{B}_{1}^2}-\frac{\mathcal{B}_2}{\mathcal{C}_2}\frac{1}{(\Delta\tau-is\beta-i\varepsilon)^2-\mathcal{B}_{2}^2}\right]\label{wightman-coC1}
\end{equation}
for $\xi=\xi'$ with $\mathcal{B}_1=\frac{2}{\alpha}\sinh^{-1}\left(\frac{nL\alpha}{2}\right),\,\mathcal{C}_1=nL\sqrt{1+\frac{1}{4}\alpha^2 n^2L^2},\,\mathcal{B}_2=\frac{2}{\alpha}\sinh^{-1}\left(\alpha\left(z_{\xi}-\frac{nL}{2}\right)\right),\,\mathcal{C}_2=(2z_{\xi}-nL)\sqrt{1+\frac{1}{4}\alpha^2 (2z_{\xi}-nL)^2}$\\
and 
\begin{equation}
G^{+}_{\beta}(x_{\xi}(\tau),x_{\xi'}(\tau'))=-\frac{1}{4\pi^{2}}\displaystyle\sum_{n=-\infty}^{\infty}\displaystyle\sum_{s=-\infty}^{\infty}\left[\frac{\mathcal{B}_3}{\mathcal{C}_3}\frac{1}{(\Delta\tau-is\beta-i\varepsilon)^2-\mathcal{B}_{3}^2}-\frac{\mathcal{B}_4}{\mathcal{C}_4}\frac{1}{(\Delta\tau-is\beta-i\varepsilon)^2-\mathcal{B}_{4}^2}\right]\label{wightman-coC2}
\end{equation}
for $\xi\neq\xi'$ with $\left\{\mathcal{B}_3=-\frac{2}{\alpha}\sinh^{-1}\left(\frac{(d+nL)\alpha}{2}\right),\,\mathcal{C}_3=-(d+nL)\sqrt{1+\frac{1}{4}\alpha^2 (d+nL)^2}\right\}\,(\text{for}\,\xi=A,\,\xi'=B),\,\left\{\mathcal{B}_3=\frac{2}{\alpha}\sinh^{-1}\left(\frac{(d-nL)\alpha}{2}\right),\,\mathcal{C}_3=(d-nL)\sqrt{1+\frac{1}{4}\alpha^2 (d-nL)^2}\right\}\,(\text{for}\,\xi=B,\,\xi'=A),\,\,\mathcal{B}_4=\frac{2}{\alpha}\times\\
\sinh^{-1}\left(\alpha
\left(z_{0}+\frac{d-nL}{2}\right)\right)$ and $\mathcal{C}_4=(2z_{0}+d-nL)\sqrt{1+\frac{1}{4}\alpha^2 (2z_{0}+d-nL)^2}$.

Inserting the above Wightman functions into eq.(s)(\ref{responserate-co}, \ref{transprobrate-co}), and performing the integrations using the contour integration technique, the upward and downward transition rates of the two-atom system submerged in the thermal bath read
\begin{align}\label{upco_C1}
\mathcal{R}^{\beta}_{\vert \psi\rangle\rightarrow \vert e_{A}e_{B}\rangle}=&\lambda^2\left\{\left(\frac{\omega_0}{2\pi}+\mathfrak{f}\left(\omega_{0},\,\alpha,\,\frac{L}{2}\right)-\cos^2\theta\,\mathfrak{h}\left(\omega_{0},\,\alpha,\,z_{0},\frac{L}{2}\right)-\sin^2\theta\,\right.\right.\nonumber\\
&\times\left.\mathfrak{m}(\omega_{0},\,\alpha,\,z_{0},\,d,\,\frac{L}{2})+\sin2\theta\,\mathfrak{n}\left(\omega_{0},\,\alpha,\,\frac{d}{2},\,\frac{L}{2}\right)- \sin2\theta\right.\nonumber\\
&\times \left.\left.\mathfrak{m}\left(\omega_{0},\,\alpha,\,z_{0},\,\frac{d}{2},\,\frac{L}{2}\right)\right)\left(\frac{1}{\exp{\omega_{0}/T}-1}\right)\right\}
\end{align}
\begin{align}\label{dwnco_C1}
\mathcal{R}^{\beta}_{\vert \psi\rangle\rightarrow \vert g_{A}g_{B}\rangle}=&\lambda^2\left\{\left(\frac{\omega_0}{2\pi}+\mathfrak{f}\left(\omega_{0},\,\alpha,\,\frac{L}{2}\right)-\cos^2\theta\,\mathfrak{h}\left(\omega_{0},\,\alpha,\,z_{0},\frac{L}{2}\right)-\sin^2\theta\,\right.\right.\nonumber\\
&\times\left.\mathfrak{m}(\omega_{0},\,\alpha,\,z_{0},\,d,\,\frac{L}{2})+\sin2\theta\,\mathfrak{n}\left(\omega_{0},\,\alpha,\,\frac{d}{2},\,\frac{L}{2}\right)- \sin2\theta\right.\nonumber\\
&\times \left.\left.\mathfrak{m}\left(\omega_{0},\,\alpha,\,z_{0},\,\frac{d}{2},\,\frac{L}{2}\right)\right)\left(1+\frac{1}{\exp{\omega_{0}/T}-1}\right)\right\}
\end{align}
where 
\begin{align}
\mathfrak{f}\left(\Delta E,\,\alpha,\,\frac{L}{2}\right)&=2\displaystyle\sum_{n=1}^{\infty}\mathfrak{g}\left(\Delta E,\,\alpha,\,\frac{nL}{2}\right)\,\\
\mathfrak{h}\left(\Delta E,\,\alpha,\,z_0\,,\frac{L}{2}\right)&=\displaystyle\sum_{n=-\infty}^{\infty}\mathfrak{g}\left(\Delta E,\,\alpha,\,z_0-\frac{nL}{2}\right)\,\\
\mathfrak{m}\left(\Delta E,\,\alpha,\,z_0\,,d,\,\frac{L}{2}\right)&=\displaystyle\sum_{n=-\infty}^{\infty}\mathfrak{g}\left(\Delta E,\,\alpha,\,z_0+d-\frac{nL}{2}\right)\,\\
\mathfrak{n}\left(\Delta E,\,\alpha,\,\frac{d}{2},\,\frac{L}{2}\right)&=\displaystyle\sum_{n=-\infty}^{\infty}\mathfrak{g}\left(\Delta E,\,\alpha,\,\frac{d-nL}{2}\right)\,\label{nfunc1}
\end{align}
where $\mathfrak{g}\left(\Delta E,\,\alpha,\,z_0\right)$ is defined as
\begin{equation}
\mathfrak{g}\left(\Delta E,\,\alpha,\,z_0\right)=\frac{\sin(\frac{2\Delta E}{\alpha}\sinh^{-1}(\alpha z_0))}{4\pi z_0\sqrt{1+\alpha^2 z_0^2}}\,.\label{gfunc1}
\end{equation}
By taking the limiting cases of the expressions eq.(s)(\ref{upco_C1}, \ref{dwnco_C1}), one can obtain the results of single mirror and free space scenarios. Taking the limit $L\rightarrow\infty$,  eq.(s)(\ref{upco_C1}, \ref{dwnco_C1}) reduce to the expression for the upward and the downward transition rates in the presence of a single reflecting boundary, respectively given by
\begin{align}
\mathcal{R}^{\beta}_{\vert \psi\rangle\rightarrow \vert e_{A}e_{B}\rangle}=&\lambda^2\Bigg\{\left(\frac{\omega_{0}}{2\pi}-\cos^2\theta\,\mathfrak{g}(\omega_{0},\,\alpha,\,z_{0})-\sin^2\theta\,\mathfrak{g}(\omega_{0},\,\alpha,\,(z_{0}+d))\right.\Bigg.\nonumber\\
&+\left.\left.\sin2\theta\left(\mathfrak{g}\left(\omega_{0},\,\alpha,\,\frac{d}{2}\right)-\mathfrak{g}\left(\omega_{0},\,\alpha,\,z_{0}+\frac{d}{2}\right)\right)\right)\left(\frac{1}{\exp{\omega_{0}/T}-1}\right)\right\}
\end{align}
\begin{align}
\mathcal{R}^{\beta}_{\vert \psi\rangle\rightarrow \vert g_{A}g_{B}\rangle}=&\lambda^2\Bigg\{\left(\frac{\omega_{0}}{2\pi}-\cos^2\theta\,\mathfrak{g}(\omega_{0},\,\alpha,\,z_{0})-\sin^2\theta\,\mathfrak{g}(\omega_{0},\,\alpha,\,(z_{0}+d))\right.\Bigg.\nonumber\\
&+\left.\left.\sin2\theta\left(\mathfrak{g}\left(\omega_{0},\,\alpha,\,\frac{d}{2}\right)-\mathfrak{g}\left(\omega_{0},\,\alpha,\,z_{0}+\frac{d}{2}\right)\right)\right)\left(1+\frac{1}{\exp{\omega_{0}/T}-1}\right)\right\}\,.
\end{align}
On the other hand, taking the limits $L\rightarrow\infty$ and $z_{0}\rightarrow\infty$ simultaneously, eq.(s)(\ref{upco_C1}, \ref{dwnco_C1}) lead to the expression for the upward and the downward transition rates in  free space given by eq.(s)(\ref{upco_emp}, \ref{dwnco_emp}).

From the above analysis of the two-atom system confined to a cavity, we  find a similarity between an instantaneously inertial observer and a coaccelerated observer in a thermal bath for both the upward and the downward transition rates. Here too it may be noted that upon taking the thermal bath temperature in the coaccelerated frame $T=\alpha/2\pi$, from eq.(s)(\ref{upco_C1}, \ref{dwnco_C1}, \ref{up_C}, \ref{dwn_C}) it follows that the transition rates of the uniformly accelerated two-atom system in the  generic entangled state seen by a coaccelerated observer and by an instantaneously inertial observer are identical inside the cavity.


\section{Conclusions}\label{sec:Con}
In this study, we have investigated the transition rates of uniformly accelerated single and entangled two-atom systems. The two-atom system is assumed to be prepared in the most generic pure entangled state. Both systems interact with the massless scalar field from the perspective of an instantaneously inertial observer and a coaccelerated observer, respectively. We have studied  the interaction between the accelerated atomic systems and the massless scalar field  in two scenarios, namely, free space and inside a cavity. We have presented two examples of the  computation of the actual
values of the transition rates using realistic system and cavity parameters.

Considering that the scalar field with which the atoms interact in the inertial frame and the coaccelerated frame, are in the vacuum state and a thermal state, respectively, it is seen that in all the above cases, both the upward  and the downward transitions  take place for the single as well as the entangled two-atom system. The upward transition is non-trivial, and from the view point of an inertial observer, takes place only due to the acceleration of the atomic systems. Our
study shows we  that the  transition rate depends on the cavity parameters, such as the length of the cavity ($L$), distance of an atom from one boundary ($z_0$), as well as other system parameters, such as the atomic acceleration ($\alpha$), the interatomic distance ($d$), and the
magnitude of initial atomic entanglement ($\theta$). 

From the analysis, it is observed that for a single atom, the upward transition rate increases with the increment of atomic acceleration and cavity length. The transition rate exhibits an oscillatory behaviour with
respect to the distance between the atom and the reflecting boundary. In case of the two-atom system, the transition rates shows some interesting features.
In this scenario, the entanglement parameter and the interatomic distance play  important roles. The transition rate shows oscillatory behaviour in the full
range of the entanglement parameter. However, considering a small magnitude
of initial entanglement, we find that in the free space, increasing the entanglement parameter enhances the upward transition and downward transition rates, whereas, in the presence of cavity it shows a completely opposite behaviour, and both transition rates get suppressed due to the increase of the entanglement parameter. In the case when the entanglement parameter has the
value $\theta=\pi/4$, we observe that both the transition rates vanish, 
indicating that no transition there occurs  from the maximally entangled initial state to any higher or lower energetic product state. Hence, the entanglement of the initial state can be preserved. From a quantum information theoretic viewpoint, this result is of significance, since preservation
of entanglement enables its
use  as resource for performing various tasks.
 
Our study further reveals  that the upward transition rate diminishes
beyond a certain level of increase of the interatomic distance. From a 
physical perspective, one may view this result to originate from the 
fact that the cooperative effects of the two atoms mediated by the field
become more and more subdued as the interatomic distance increases beyond
a point. On the other hand, the efect due to the distance of an atom from the
boundary has a more subtle manifestation. We find that the upward transition rate increases when we increase the distance between any one of the atoms and one boundary, and takes the maximum value when the distance between both atoms to their closest boundaries are equal.  Apart from this, the behaviour of the transition rate with respect to atomic acceleration and the cavity length is quite similar to the single atom case.
 
From our extensive study of the transition rates of the single and two-atom systems in an inertial and a coaccelerated frame, we observe that if the temperature of the thermal bath in the coaccelerated frame is taken to be the
same as the Unruh temperature, then the transition rates for the upward and the downward transitions in the two frames coincide exactly  with each other  even inside the cavity, making it  completely consistent with the Fulling-Davies-Unruh effect. Therefore, from the present study,   the equivalence between the effect of uniform acceleration and the effect of thermal bath 
is clearly manifested for the single as well as the entangled two-atom cases in free space and in the presence of reflecting boundaries, as well. The finding is intriguing for the cavity case since the physics changes quite a bit inside a cavity, and moreover such a set-up is  experimentally implementable \cite{Lochan2020, Lochan2022, Vetsch2010, Goban2012}.
\section*{Acknowledgement}
AM and ASM acknowledges support from project no. DST/ICPS/QuEST/2019/Q79 of the Department of Science and Technology (DST), Government of India.
\begin{appendices}
\section{Calculation of some useful integrals used in the text}\label{Appendix:A}
In this Appendix, for the sake of completeness, we provide a detailed calculation of the following integrals
\begin{align}
\mathcal{I}_{1}&=
 -\frac{\alpha^2}{16\pi^2}\int_{-\infty}^{+\infty} d(\Delta\tau) \,e^{-i\Delta E\Delta\tau}\frac{1}{\sinh^2\left[\frac{1}{2}(\alpha\Delta\tau-i\varepsilon)\right]}\label{A1}\\
\mathcal{I}_{2}&= -\frac{\alpha^2}{16\pi^2}\int_{-\infty}^{+\infty} d(\Delta\tau) \,e^{-i\Delta E\Delta\tau}\frac{1}{\sinh^2\left[\frac{1}{2}(\alpha\Delta\tau-i\varepsilon)\right]-d^2\alpha^2}\,.\label{A2}
\end{align}
To solve the integral given in eq.\eqref{A1}, we first consider some dimensionless parameters such as $\alpha\Delta\tau=\sigma,\, \Delta E/\alpha=\xi$, and use the series representation \cite{Gradshteyn:1702455}
\begin{equation}
\csch^2\left[\frac{1}{2}\left(\sigma -i\varepsilon\right)\right]=\displaystyle\sum_{k=-\infty}^{\infty}\frac{4}{(\sigma -i\varepsilon -2i\pi k)^2}\,.\label{csch2-series}
\end{equation}
Eq.\eqref{A1} now becomes
\begin{equation}\label{A4}
\mathcal{I}_1=-\frac{\alpha}{4\pi^2}\displaystyle\sum_{k=-\infty}^{\infty}\int_{-\infty}^{+\infty} d\sigma\frac{e^{-i\xi\sigma}}{(\sigma -i\varepsilon -2i\pi k)^2}=-\frac{\alpha}{4\pi^2}\Bigg[\mathcal{I}_0
+\displaystyle\sum_{k=1}^{\infty}\mathcal{I}_k\Bigg]
\end{equation}
where $\mathcal{I}_0$ and $\mathcal{I}_k$ are given by the integrals
\begin{align}
\mathcal{I}_{0}&=\int_{-\infty}^{+\infty}d\sigma\,\frac{e^{-i\xi\sigma}}{(\sigma-i\varepsilon)^2}\label{A5}\\
\mathcal{I}_{k}&=\int_{-\infty}^{+\infty}d\sigma\,e^{-i\xi\sigma}\left\{\frac{1}{(\sigma-2i\pi k)^2}+\frac{1}{(\sigma+2i\pi k)^2}\right\}\,\,.\label{A6}
\end{align}
\begin{figure}[H]
\centering
\includegraphics[scale=0.55]{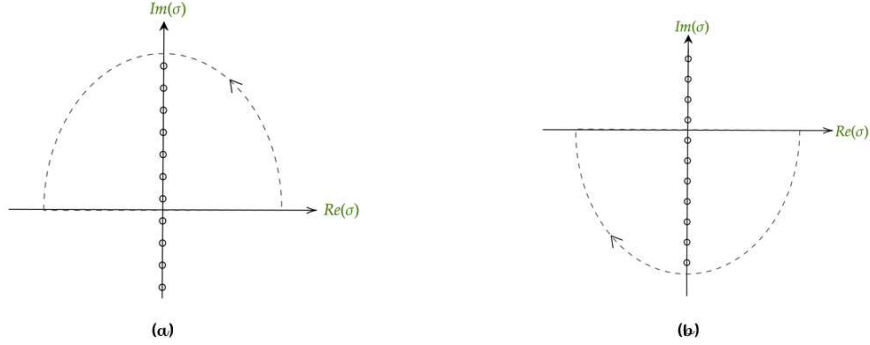}
\caption{The contour of the integral \eqref{A1}for (a) $\Delta E<0$ and (b) $\Delta E>0$.}\label{fig:Pole1}
\end{figure}
\noindent Considering the analytic continuation of the above integrands in the complex plane of $\sigma$, for eq.\eqref{A5} we get a second order pole at $\sigma=i\varepsilon$ and for eq.\eqref{A6} we find a set of second order poles at $\sigma=-2i\pi k$ with $k=\pm1,\pm2,\ldots$, lying on the imaginary axis of $\sigma$. For $\xi<0$ or $\Delta E<0$, we close the contour in the upper half complex plane in Figure \ref{fig:Pole1}. Using Jordon's lemma, we observe that the integration along the half circle is zero. Therefore, applying the Cauchy residue theorem in the integrals \eqref{A5} and \eqref{A6} and using it in eq.\eqref{A4}, we get
\begin{equation}\label{A7}
\mathcal{I}_1(\Delta E<0)=-\frac{\alpha}{4\pi^2}\Bigg[2\pi\xi
+2\pi\xi\displaystyle\sum_{k=1}^{\infty} e^{2\pi k\xi}\Bigg]=\frac{\alpha\vert\xi\vert}{2\pi}\left[1+\frac{1}{e^{2\pi\vert\xi\vert}-1}\right]=\frac{\vert\Delta E\vert}{2\pi}\left[1+\frac{1}{e^{2\pi\vert\Delta E\vert/\alpha}-1}\right]\,.
\end{equation}
Similarly, for $\Delta E>0$, closing the contour in the lower half complex plane and following the previous steps, we get
\begin{equation}\label{A8}
\mathcal{I}_1(\Delta E>0)=\frac{\Delta E}{2\pi}\frac{1}{e^{2\pi\Delta E/\alpha}-1}\,.
\end{equation}
Therefore, combining the above results \eqref{A7} and \eqref{A8}, we get
\begin{equation}\label{A9}
\mathcal{I}_{1}=\theta\,(-\Delta E)\frac{\vert \Delta E\vert}{2\pi}\left(1+\frac{1}{\exp(2\pi\vert \Delta E\vert/\alpha)-1}\right)+\theta\,(\Delta E)\frac{\Delta E}{2\pi}\left(\frac{1}{\exp(2\pi\Delta E/\alpha)-1}\right)
\end{equation}
where $\theta(\Delta E)$ is the Heaviside step function.\\
\begin{figure}[H]
\centering
\includegraphics[scale=0.5]{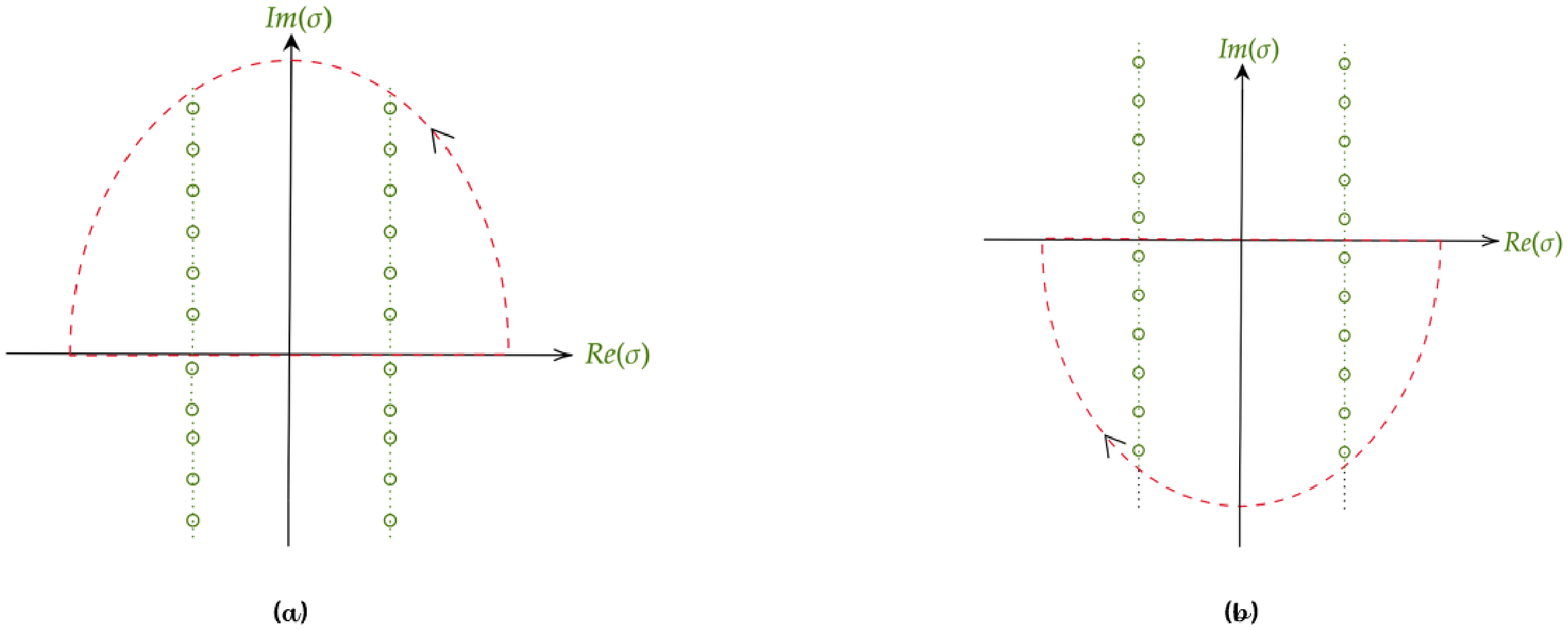}
\caption{The contour of the integral \eqref{A2} for (a) $\Delta E<0$ and (b) $\Delta E>0$.}\label{fig:Pole2}
\end{figure}
\noindent To evaluate the integral in eq.\eqref{A2}, first we consider the analytic continuation of the integrand in the complex plane $\sigma$. Eq.\eqref{A2} then becomes
\begin{equation}\label{A10}
\mathcal{I}_{2}= -\frac{\alpha}{16\pi^2}\oint_{C} d\sigma \,e^{-i\xi\sigma}\frac{1}{\sinh^2\left[\frac{1}{2}(\sigma-i\varepsilon)\right]-d^2\alpha^2}
\end{equation}
where $C$ is the contour in Figure \ref{fig:Pole2}.
From the integrand we find that there exists two types of first order poles $\sigma^{+}=i\varepsilon+2in\pi+2\sinh^{-1}(d\alpha)$ and $\sigma^{-}=i\varepsilon+2in\pi-2\sinh^{-1}(d\alpha)$ where $n=0,\pm1,\pm2,\ldots$, lying in the upper and lower half of the complex plane.  Therefore, closing the contour in the upper half complex plane for $\Delta E<0$, we find that both the poles will contribute for all positive integer values of $n$ including $n=0$.\\
 Now applying the residue theorem and taking the limit $\varepsilon\rightarrow0$, the residues become
\begin{equation}\label{A11}
R_{1}(n)=\frac{\exp{-i\frac{\Delta E}{\alpha}(2in\pi+2\sinh^{-1}(d\alpha))}}{\alpha d\sqrt{1+d^2\alpha^2}}
\end{equation}
and
\begin{equation}\label{A11a}
R_{2}(n)=-\frac{\exp{-i\frac{\Delta E}{\alpha}(2in\pi-2\sinh^{-1}(d\alpha))}}{\alpha d\sqrt{1+d^2\alpha^2}}\,.
\end{equation}
Applying Jordon's lemma, we already observe that the integration along the half circle is zero, therefore we get
\begin{align}\label{A12}
\mathcal{I}_{2}(\Delta E<0)&=-\frac{\alpha}{16\pi^2}\displaystyle\sum_{n=0}^{\infty}2\pi i\Big[R_{1}(n)+R_{2}(n)\Big]\nonumber\\
&=\frac{\sin(\frac{2\vert\Delta E\vert}{\alpha}\sinh^{-1}(d\alpha))}{4\pi d\sqrt{1+d^2\alpha^2}}\left(1+\frac{1}{e^{2\pi\vert\Delta E\vert/\alpha}-1}\right)\,.
\end{align}
Similarly, for $\Delta E>0$, closing the contour in the lower half complex plane we find that both the poles will contribute for all negative integer values of $n$.\\
Now following the steps of the case $\Delta E<0$, we get
\begin{equation}\label{A13}
\mathcal{I}_2(\Delta E>0)=\frac{\sin(\frac{2\Delta E}{\alpha}\sinh^{-1}(d\alpha))}{4\pi d\sqrt{1+d^2\alpha^2}}\left(\frac{1}{e^{2\pi\Delta E/\alpha}-1}\right)\,.
\end{equation}
Therefore, combining the above results \eqref{A12} and \eqref{A13}, we get
\begin{equation}\label{A14}
\mathcal{I}_{2}=\theta(-\Delta E)\frac{\sin(\frac{2\vert\Delta E\vert}{\alpha}\sinh^{-1}(d\alpha))}{4\pi d\sqrt{1+d^2\alpha^2}}\left(1+\frac{1}{e^{2\pi\vert \Delta E\vert/\alpha}-1}\right)+\theta(\Delta E)\frac{\sin(\frac{2\Delta E}{\alpha}\sinh^{-1}(d\alpha))}{4\pi d\sqrt{1+d^2\alpha^2}}\left(\frac{1}{e^{2\pi\Delta E/\alpha}-1}\right)\,.
\end{equation}
\end{appendices}
\bibliographystyle{hephys}
\bibliography{Reference.bib}
\end{document}